\documentclass[journal]{IEEEtran}

\ifCLASSINFOpdf
\else
   \usepackage[dvips]{graphicx}
\fi

\usepackage{url}
\usepackage{graphicx,cite,epsfig,amssymb,amsmath,color}
\usepackage{subfigure}
\usepackage{float}
\usepackage{algorithm}
\usepackage{algpseudocode}
\usepackage [OT1] {fontenc}
\usepackage{cuted}
\usepackage{stfloats}
\usepackage{comment}
\usepackage{ragged2e}
\usepackage{booktabs} % 导入三线表需要的宏包

\def\BibTeX{{\rm B\kern-.05em{\sc i\kern-.025em b}\kern-.08em
    T\kern-.1667em\lower.7ex\hbox{E}\kern-.125emX}}
\begin{document}

\title{Achievable Rate Region and Path-Based Beamforming  for Multi-User Single-Carrier\\
 Delay Alignment Modulation}

\author{Xingwei Wang, Haiquan Lu, Yong Zeng, \emph{Senior Member}, \emph{IEEE},    Xiaoli Xu, \emph{Member}, \emph{IEEE}, \\ and Jie Xu, \emph{Senior Member}, \emph{IEEE} 
\thanks{This work was supported by the National Key R\&D Program of China with Grant number 2019YFB1803400 and also by the Natural Science Foundation of China under Grant 62071114. Part of this work will be presented at the IEEE Globecom 2023 \cite{MU-DAM}. (Corresponding author: Yong Zeng.)}
\thanks{Xingwei Wang, Haiquan Lu, and Yong Zeng are with the National Mobile Communications Research Laboratory, Southeast University, Nanjing 210096, China. Haiquan Lu and Yong Zeng  are also with the Purple Mountain Laboratories, Nanjing 211111, China. Xiaoli Xu is with the School of Information Science and Engineering, Southeast University, Nanjing
210096, China. Jie Xu is with the School of Science and Engineering and Future Network of Intelligence Institute, The Chinese University of Hong Kong (Shenzhen), Shenzhen 518172, China. (e-mail: \{xingwei-wang, haiquanlu,   yong\_zeng, xiaolixu\}@seu.edu.cn, xujie@cuhk.edu.cn)}}

\maketitle

\begin{abstract}
Delay alignment modulation (DAM) is a novel wideband transmission technique for millimeter wave (mmWave) massive multiple-input multiple-output (MIMO) systems, which exploits the high spatial resolution and multi-path sparsity to mitigate inter-symbol interference (ISI), without relying on  channel equalization or multi-carrier transmission. In particular,  DAM leverages the {\it delay pre-compensation} and {\it path-based beamforming} to effectively align the multi-path components, thus achieving the constructive multi-path combination for eliminating  the ISI while preserving the multi-path power gain. Different from the existing works only considering single-user DAM, this paper investigates  the DAM technique for multi-user mmWave massive MIMO communication. First, we consider the asymptotic regime when the number of antennas $M_t$ at  base station (BS) is sufficiently large. It is shown that by employing the simple delay pre-compensation and per-path-based maximal ratio transmission (MRT) beamforming, the single-carrier DAM is able to perfectly eliminate both ISI and inter-user interference (IUI). Next, we consider the general scenario with $M_t$ being finite. In this scenario, we characterize the achievable rate region of the multi-user DAM system by finding its Pareto boundary.  Specifically, we formulate a rate-profile-constrained sum rate maximization problem by optimizing the per-path-based beamforming, which is optimally solved  via the second-order cone programming (SOCP). Furthermore, we present three low-complexity per-path-based   beamforming strategies based on the  MRT, zero-forcing (ZF), and regularized zero-forcing (RZF) principles, respectively, based on which the achievable sum rates are studied.  Finally, we provide simulation results to demonstrate the performance of our proposed strategies as compared to two benchmark schemes based on the strongest-path-based beamforming and the  prevalent orthogonal frequency division multiplexing (OFDM), respectively. It is shown that DAM achieves higher spectral efficiency and/or lower peak-to-average-ratio (PAPR), for systems with high spatial resolution and multi-path diversity. 

%for systems with high spatial resolution and multi-path sparsity, DAM outperforms the two benchmarking schemes in terms of spectral efficiency and/or peak-to-average-power ratio (PAPR), thanks to its capability  of eliminating ISI while reaping the multi-path power gain with single-carrier communication  and  small guard interval overhead.
\end{abstract}
\

\IEEEpeerreviewmaketitle

%\begin{comment}
\section{Introduction}\label{introduction}

%latex 参考文献合并\cite{DAM-CSI,DAM-OFDM,DAM-ISAC,DAM-ISAC-other,DAM-IRSs}
%展开来说明，对于单用户极限情况，以及DAM不但可以与OFDM兼容，还可以和IRS兼容，以及ISAC兼容等等 and peak-to-average-power ratio (PAPR)
Recently, delay alignment modulation (DAM)  was proposed as a novel technique to tackle the  inter-symbol interference (ISI) issue in wideband communication systems, without relying  on  channel equalization or multi-carrier transmissions \cite{DAM}. DAM is particularly appealing for systems with large antenna arrays \cite{spatial, Lu}  and high-frequency bands such as millimeter wave (mmWave) \cite{mmWave} and Terahertz (THz) systems,  which have the super spatial resolution  and  multi-path sparsity. In particular, by  manipulating the channel delay spread via {\it delay pre-compensation} and {\it path-based beamforming}, DAM enables all multi-path signal components to reach the receiver concurrently and constructively, thus eliminating the detrimental ISI while preserving the multi-path channel power gains. This renders DAM resilient to time-dispersive channels for efficient single- and multi-carrier transmissions. There have been some preliminary works on DAM in the literature \cite{DAM-CSI,DAM-OFDM,DAM-ISAC,DAM-ISAC-other,DAM-IRSs}. For example, by combining DAM with orthogonal frequency division multiplexing (OFDM), a novel DAM-OFDM scheme was proposed in \cite{DAM-OFDM}, which provides a unified framework to achieve ISI-free communication using single- or multi-carrier transmissions by flexibly manipulating the channel delay spread and the number of sub-carriers. Furthermore, DAM was exploited  for  integrated sensing and communication (ISAC) and multi-intelligent reflecting surfaces (IRSs) aided communication  in \cite{DAM-ISAC}, \cite{DAM-ISAC-other} and \cite{DAM-IRSs}, respectively.

DAM is significantly  different from existing ISI-mitigation techniques,  such as channel equalization, spread spectrum, and multi-carrier transmission. First, channel equalization can be generally classified into time- and frequency-domain equalization, respectively. For time-domain equalization, a particular time reversal (TR) for single-carrier transmission was proposed in \cite{TR1} by treating the multi-path channel as an intrinsic matched filter \cite{TR1,TR3}, in which the ISI issue is addressed by compromising the communication rate due to the use of  rate back-off technique. In \cite{CPshort1} and \cite{CPshort2}, channel shortening technique was proposed by applying a time-domain finite impulse response (FIR) filter, in which the effective channel impulse response is shortened to avoid long cyclic prefix (CP). However, the time-domain equalization needs a large number of taps for systems with large channel delay spread. Besides  time-domain equalization, frequency-domain equalization reduces the signal processing complexity by converting  the signal to the frequency-domain via  discrete Fourier transform (DFT) \cite{CPshort2}, \cite{TDandFD}. By contrast, DAM can completely eliminate ISI via spatial-delay processing.  Spread spectrum techniques like RAKE receiver \cite{goldsmith} require spread spectrum codes with good auto/cross-correlation and bandwidth expansion, i.e., using bandwidth much larger than  necessary for data transmission, while DAM avoids these issues. In addition, OFDM and OTFS are two widely adopted multi-carrier  communication technologies. OFDM suffers from  practical issues like sensitivity to carrier frequency offset (CFO), high peak-to-average-power ratio (PAPR), and the severe out-of-band (OOB) emission \cite{goldsmith}.  Moreover, orthogonal time-frequency space (OTFS) modulation was recently proposed by modulating information  in the delay-Doppler domain, which  shows  superior performance to OFDM in high mobility scenarios \cite{OTFSbook}. However, the multi-carrier OTFS  technique has high  complexity when extending to massive MIMO systems \cite{OTFS-overview}. In addition, OTFS in general needs  to perform signal detection across the entire OTFS frame, thus leading to a  high detection delay.  By contrast, as a spatial-delay processing technique, DAM is expected to overcome  the aforementioned issues.
For example, compared to OFDM and OTFS, single-carrier DAM enables the instant symbol-by-symbol signal detection at the receiver \cite{DAM-IRSs}. 
%The reducing channel delay spread for DAM technique allows the RAKE receiver perform better\cite{8}. 

DAM is also different from various delay pre-compensation techniques in \cite{TDL, SPL, lens, time-deay}. For example, unlike DAM, \cite{TDL} and \cite{SPL} did not exploit the  high spatial resolution or multi-path sparsity of mmWave massive MIMO systems to  eliminate ISI. In addition, for mmWave MIMO systems, \cite{lens} and  \cite{time-deay}  were designed  for the special lens MIMO or hybrid beamforming architectures, respectively. 
Besides, equalization-free single-carrier modulation was also advocated in \cite{BeamA} and \cite{little-no-e}, where beam alignment along the single dominant path was applied to transform the time-variant frequency-selective channel into time-invariant frequency-flat channel \cite{BeamA}. Furthermore, the measurement campaign in \cite{little-no-e} showed that little or no equalization is needed with such a beam alignment scheme. However, both \cite{BeamA} and \cite{little-no-e} only treat the single dominant path as the desired signal and all other multi-path signal as noise, while DAM makes full use of all multipath channel components and hence a better performance is expected, as demonstrated in the simulation results later.

Note that existing works on  DAM  \cite{DAM}, \cite{DAM-CSI,DAM-ISAC-other,DAM-OFDM,DAM-ISAC,DAM-IRSs} only consider the single-user scenario. In this paper, we study the more general multi-user  DAM  systems. In contrast to single-user DAM that only addresses the ISI issue, multi-user DAM design needs to handle both the ISI and inter-user interference (IUI). The main contributions of this paper are summarized as follows:

\begin{itemize}
\item  First, to gain essential insights, we provide an asymptotic analysis by assuming that the number of BS antennas $M_t$ is sufficiently large.  In this case, we show that with the low-complexity delay pre-compensation and per-path-based MRT  beamforming,  the multi-user time-dispersive broadcast channel with ISI and IUI can be transformed into ISI-free and IUI-free additive white Gaussian noise (AWGN) channels.

%开始阅读！！！！！！！！！！！！！！，要不要增加benchmark
%{\color{red}Revise accordingly based on the abstract. Give more details.}
\item  Second, for the general scenario with a finite number of BS antennas, we characterize the achievable rate region of the multi-user DAM system by finding its Pareto boundary. The rate-profile-constrained sum rate maximization problem is formulated  by optimizing the per-path-based beamforming, which is optimally solved  via second-order cone programming (SOCP).

%{\color{red}this part considers the sum rate.... Revise the presentation accordingly} 
\item  Third, we present three low-complexity  beamforming strategies in a per-path basis, namely MRT, zero-forcing (ZF), and regularized zero-forcing (RZF), respectively. Specifically, low complexity per-path-based MRT beamforming for multi-user DAM with tolerable  ISI and IUI is firstly developed. Then, when $M_t\geq L_{\mathrm{tot}}$, both the ISI and IUI are completely eliminated with the per-path-based ZF beamforming, and the optimal  power allocation for sum rate maximization  is obtained. Furthermore, the more general per-path-based RZF beamforming with tolerable  ISI and IUI is developed to balance the noise enhancement in the ZF beamforming and the interference limitation in the MRT beamforming.
%{\color{red}How about the single-carrier strongest path scheme?}  
\item Last,  we compare the proposed multi-user single-carrier DAM designs with two benchmarking schemes, i.e., the
alternative single-carrier ISI suppression technique that only uses the strongest channel path of each user and the
prevalent OFDM scheme.  Extensive  simulation results are provided to  demonstrate the advantages of single-carrier DAM over the benchmarking schemes, in terms of spectral efficiency and/or  PAPR. 
\end{itemize}

The rest of this paper is organized as follows. Section II presents the system model and provides the asymptotic analysis when $M_t$ is sufficiently large. Section III characterizes the  achievable rate region for  multi-user DAM. Section IV studies the  sum rate for multi-user DAM.  Section V compares the multi-user DAM against the benchmarking schemes of strongest-path-based beamforming  and OFDM. Section VI presents numerical results. Finally, we conclude the paper in Section VII.
 
{\it Notations}: Scalars are denoted by italic letters. Vectors and matrices are denoted by boldface lower- and upper-case letters, respectively. For a matrix $\boldsymbol{\bf{A}} \in \mathbb{C}^{M \times N}$, $\boldsymbol{\bf{A}}^T, \boldsymbol{\bf{A}}^H, \boldsymbol{\bf{A}}^{\dagger}$,  $\|\boldsymbol{\bf{A}}\|_F$, $\Im(\boldsymbol{\bf{A}})$, and $\Re(\boldsymbol{\bf{A}})$  denote its transpose, Hermitian transpose, pseudo-inverse,  Frobenius norm, imaginary part, and real part, respectively.  For a vector $\boldsymbol{\bf{a}}$, $\|\boldsymbol{\bf{a}}\|$ denotes the  $l_2$-norm.
 $\mathbb{E}[\cdot]$, $\delta[\cdot]$, $\max [\cdot] $, and $\min [\cdot]$ denote the expectation,  Dirac-delta impulse function, maximum and minimum operators, respectively. $*$, and $\mathcal{CN}(\boldsymbol{\bf{x}},\boldsymbol{\bf{\Sigma}})$  denote linear convolution, and circularly symmetric complex Gaussian (CSCG)  distribution of a random vector with mean vector $\boldsymbol{\bf{x}}$ and covariance matrix $\boldsymbol{\bf{\Sigma}}$, respectively,  and $(a)^+\triangleq \max(a,0)$.

%\end{comment}
\begin{figure}
    \centering
     %\vspace{-0.8cm}  % 调整与上文的间距
    \includegraphics[scale=0.35]{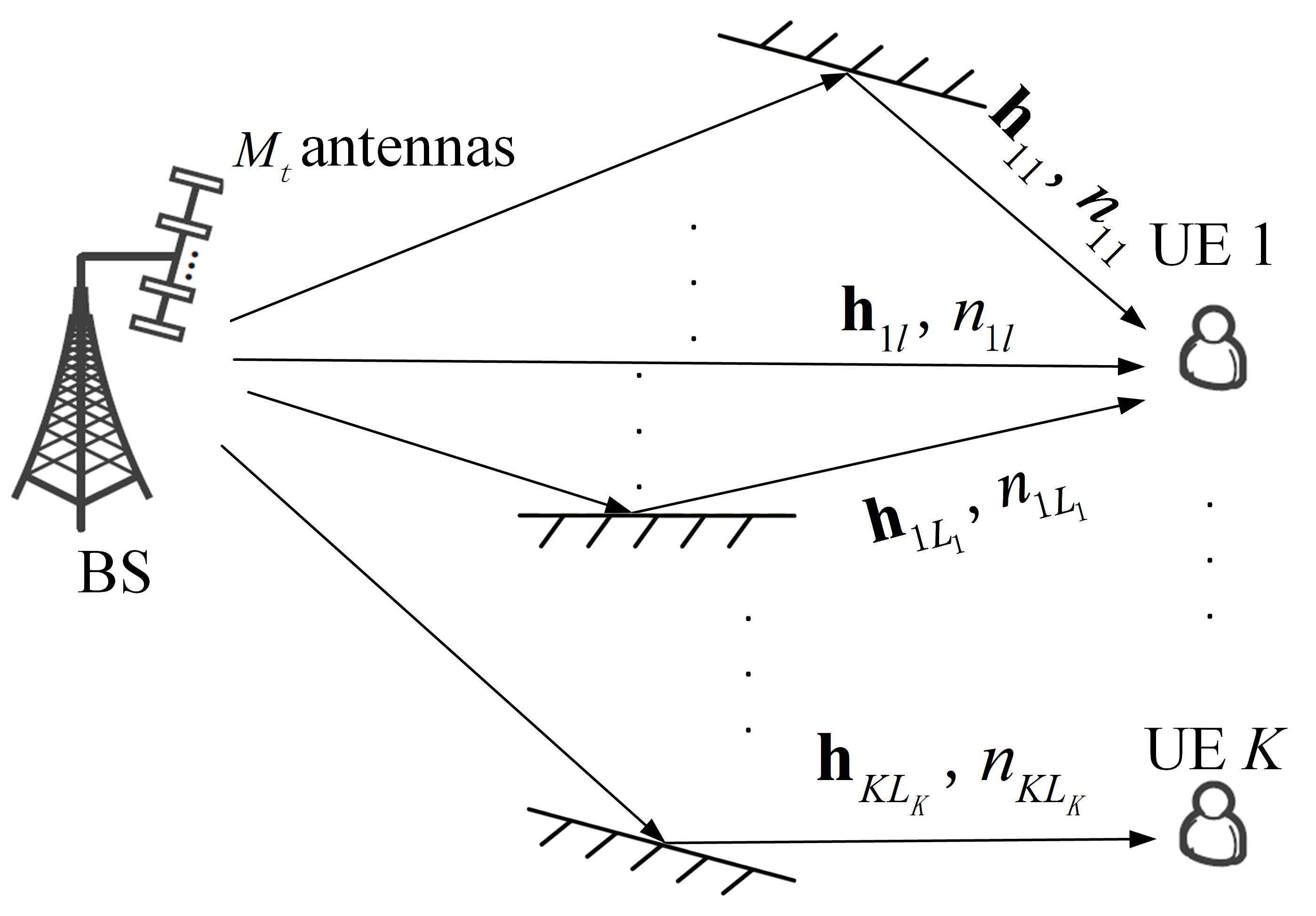}
    \caption{A multi-user mmWave massive MIMO communication system in time-dispersive channels.} \label{fig1}
    %\vspace{-0.7cm}
\end{figure}

\section{System Model and Asymptotic Analysis}\label{sec1}

As shown in Fig. \ref{fig1}, we consider a multi-user mmWave massive MIMO communication system, where the BS equipped with $M_t \gg 1$ antennas serves $K$ single-antenna user equipments (UEs). In time-dispersive  multi-path environment, the baseband discrete-time channel impulse response of UE $k$ within one channel coherence block is expressed as
\setlength\abovedisplayskip{2.5pt}
\setlength\belowdisplayskip{2.5pt}
\begin{equation}
\boldsymbol{\bf{h}}_k[n]=\sum_{l=1}^{L_k} \boldsymbol{\bf{h}}_{kl} \delta[n-n_{kl}], \label{1}
\end{equation}
where  $\boldsymbol{\bf{h}}_{kl} \in \mathbb{C}^{M_{t} \times 1}$ denotes the channel coefficient vector for the $l$th multi-path of UE $k$, $L_k$ is the number of temporal-resolvable multi-paths of UE $k$, and $n_{kl}$ denotes its discretized  delay, with  $n_{kl}\neq n_{kl'}, \forall l'\neq l$. Let $n_{k, \text{max}}=\max\limits_{1\leq l\leq L_k}n_{kl}$ and $n_{k, \text{min}}=\min\limits_{1\leq l\leq L_k}n_{kl}$  denote the maximum  and minimum delays of UE $k$, respectively. 

\begin{figure}
\centering
	\includegraphics[scale=.37]{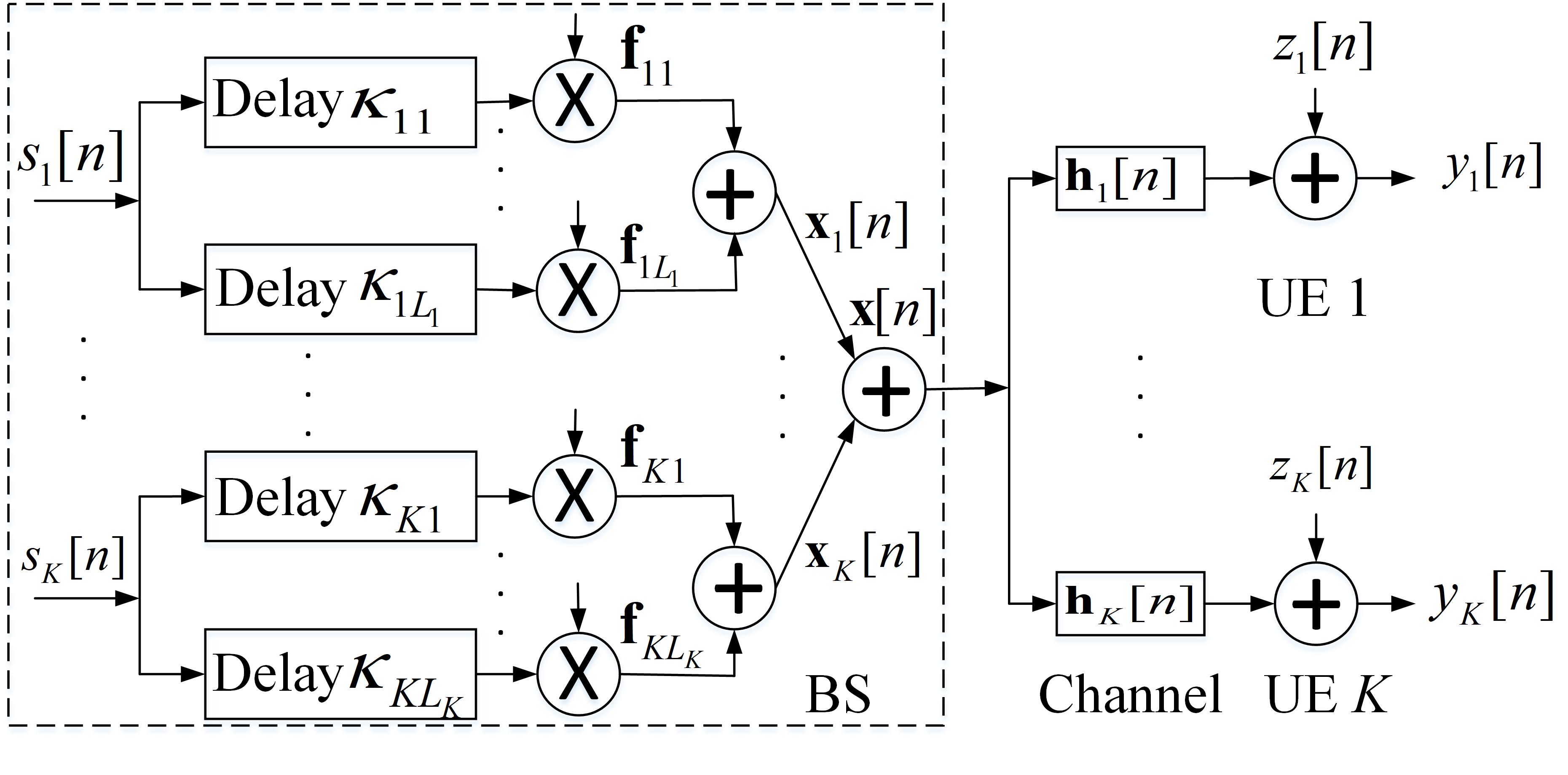}
\caption{Block diagram  for multi-user single-carrier DAM.}\label{fig2}
 %\vspace{-3ex} %%%%
\end{figure}

Let $s_k[n]$ denote the independent and identically distributed (i.i.d.) information-bearing symbols of UE $k$ with normalized power $\mathbb{E}[|s_k[n]|^2]=1$.
By extending the DAM technique proposed in \cite{DAM} to  multi-user systems, the transmitted signal by the BS for multi-user single-carrier  DAM is
\begin{equation}
\boldsymbol{\bf{x}}[n]=\sum_{k=1}^{K}\sum_{l'=1}^{L_k}\boldsymbol{\bf{f}}_{kl'} s_{k}[n-\kappa_{kl'}], \label{2}
\end{equation}
where $\boldsymbol{\bf{f}}_{kl'} \in \mathbb{C}^{M_t \times 1}$ denotes the per-path-based transmit beamforming vector associated with multi-path $l'$ of UE $k$, and $\kappa_{kl'}$ is the deliberately introduced delay pre-compensation, with $\kappa_{kl'}=n_{k, \text{max}}-n_{kl'} \geq 0, \forall l'=1,...,L_{k}$. Since $\{n_{kl'},\forall k,l'\}$ are distinct for different paths $l'$, we have $\kappa_{kl'}\neq \kappa_{kl}$,  $\forall l'\neq l$. The block diagram  of  multi-user single-carrier DAM is illustrated in Fig. \ref{fig2}. By using the fact that $s_{k}[n]$ is  independent across different $n$ and $k$ as well as  $\kappa_{kl'}\neq \kappa_{kl}, \forall l\neq l'$, the transmit power of the BS is
\begin{equation}
\mathbb{E}[\|\boldsymbol{\bf{x}}[n]\|^2]=\begin{matrix}\sum_{k=1}^K \end{matrix}\begin{matrix}\sum_{l'=1}^{L_{k}}\end{matrix}  \|\boldsymbol{\bf{f}}_{kl'}\|^2 \leq P, \label{3}
\end{equation}
where $P$ is the maximum allowable transmit power.

Based on \eqref{1} and \eqref{2}, the received signal at UE $k$ for multi-user DAM system  is
\begin{align}
&y_k[n]=\boldsymbol{\bf{h}}_k^H[n]*\boldsymbol{\bf{x}}[n]+ z_k[n]\label{4} \\  \notag
&=\underbrace{\left(\begin{matrix}\sum_{l=1}^{L_k}\end{matrix}\boldsymbol{\bf{h}}_{kl}^H\boldsymbol{\bf{f}}_{kl}\right)  s_k[n-n_{k, \text{max}}]}_{\text{Desired signal}} \\  \notag
                 & +\underbrace{\begin{matrix}\sum_{l=1}^{L_k} \end{matrix}\begin{matrix}\sum_{l'\ne l}^{L_k}\end{matrix} \boldsymbol{\bf{h}}_{kl}^H \boldsymbol{\bf{f}}_{kl'} s_k[n-n_{k, \text{max}}-n_{kl}+n_{kl'}] }_{\text{ISI}} \\  \notag
                  & +\underbrace{\sum_{l=1}^{L_k}\sum_{k'\ne k}^{K}\sum_{l'=1}^{L_{k'}}\boldsymbol{\bf{h}}_{kl}^H  \boldsymbol{\bf{f}}_{k'l'}s_{k'}[n-n_{k', \text{max}}-n_{kl}+n_{k'l'}]}_{\text{IUI}} +z_k[n],
\end{align}
where  $z_k[n]\sim \mathcal{CN}(0,\sigma^2)$ denotes  the  AWGN at UE $k$. It is observed that when UE $k$ performs the simple single-tap detection for signal with delay $n_{k, \text{max}}$, then the first term is the desired signal, and the second and third terms are the ISI and IUI, respectively.  

%

%\subsection{Asymptotic Analysis}
To show the benefit of DAM in multi-user mmWave massive MIMO, we first  provide the asymptotic analysis by considering  that  the number of BS antennas $M_t$ is much larger than the total number of temporal-resolvable  multi-paths $L_{\mathrm{tot}}=\sum_{k=1}^K L_{k}$. We will show that in this special scenario,  both the ISI and IUI in \eqref{4} asymptotically vanish  with the simple  per-path-based MRT beamforming. 

First, we consider the correlation property of the channel vectors $\boldsymbol{\bf{h}}_{kl}, k=1,...,K, l=1,...,L_k$.  Based on the asymptotic analysis in   \cite{DAM-OFDM},  it was shown that as long as the multi-paths correspond to distinct angle-of-departures (AoDs), the channel vectors of different multi-path delays are asymptotically orthogonal when $M_t \gg L_{\mathrm{tot}}$, i.e., 
\begin{equation}
\lim_{M_t \to \infty}\frac{|\boldsymbol{\bf{h}}_{kl}^H\boldsymbol{\bf{h}}_{k'l'}|}{\|\boldsymbol{\bf{h}}_{kl}\|~\|\boldsymbol{\bf{h}}_{k'l'}\|} \to 0, \forall k'\ne k, \text{or}~ l'\ne l.\label{6}
\end{equation}

By exploiting the above asymptotically orthogonal property, the ISI and IUI in \eqref{4} for the single-carrier DAM system can be eliminated by the  low-complexity per-path-based MRT beamforming. With MRT, the per-path-based beamforming vectors in \eqref{2} are set as 
\begin{equation} 
\boldsymbol{\bf{f}}_{kl'}=\xi_k\sqrt{p_k}\boldsymbol{\bf{h}}_{kl'}, k=1,...,K, l'=1,...L_k, \label{7112} 
\end{equation}
where $\xi_k=1/\sqrt{\sum_{j=1}^{L_k}\|\boldsymbol{\bf{h}}_{kj}\|^2}$ is the normalization factor and $p_k$ denotes the transmit  power allocated to UE $k$. In this case, by scaling the signal in \eqref{4} with $\xi_k$, we have 
\begin{align}
\xi_k y_k[n]&=\sqrt{p_k} s_k[n-n_{k, \text{max}}]  +\sqrt{p_k}\sum_{l=1}^{L_k} \sum_{l'\ne l}^{L_k} \frac{\boldsymbol{\bf{h}}_{kl}^H \boldsymbol{\bf{h}}_{kl'}}{\sum_{j=1}^{L_k}\|\boldsymbol{\bf{h}}_{kj}\|^2}\label{7} \nonumber \\  
                  &\times s_k[n-n_{k, \text{max}}-n_{kl}+n_{kl'}] \nonumber \\ 
                  & +\sum_{l=1}^{L_k}\sum_{k'\ne k}^{K}\sum_{l'=1}^{L_{k'}} \frac{\sqrt{p_{k'}}\boldsymbol{\bf{h}}_{kl}^H \boldsymbol{\bf{h}}_{k'l'}}{\sqrt{\sum_{j=1}^{L_k}\|\boldsymbol{\bf{h}}_{kj}\|^2}~ \sqrt{\sum_{j=1}^{L_{k'}}\|\boldsymbol{\bf{h}}_{k'j}\|^2}}
              \nonumber \\ 
                  &\times s_{k'}[n-n_{k', \text{max}}-n_{kl}+n_{k'l'}] +\xi_k z_k[n].
\end{align}
Thanks to the asymptotically orthogonal property in \eqref{6}, when $M_t \gg L_{\mathrm{tot}}$, we have 
\begin{equation}
\frac{|\boldsymbol{\bf{h}}_{kl}^H \boldsymbol{\bf{h}}_{kl'}|}{\sum_{j=1}^{L_k}\|\boldsymbol{\bf{h}}_{kj}\|^2} \leq \frac{|\boldsymbol{\bf{h}}_{kl}^H \boldsymbol{\bf{h}}_{kl'}|}{\|\boldsymbol{\bf{h}}_{kl}\|^2+\|\boldsymbol{\bf{h}}_{kl'}\|^2}\leq \frac{|\boldsymbol{\bf{h}}_{kl}^H \boldsymbol{\bf{h}}_{kl'}|}{\|\boldsymbol{\bf{h}}_{kl}\|~\|\boldsymbol{\bf{h}}_{kl'}\|}\to 0, \label{8}
\end{equation}
and
\begin{equation}
\begin{split}
&\frac{|\boldsymbol{\bf{h}}_{kl}^H \boldsymbol{\bf{h}}_{k'l'}|}{\sqrt{(\sum_{j=1}^{L_k}\|\boldsymbol{\bf{h}}_{kj}\|^2) (\sum_{j=1}^{L_{k'}}\|\boldsymbol{\bf{h}}_{k'j}\|^2)}} \label{9} \\  
&\leq \frac{|\boldsymbol{\bf{h}}_{kl}^H \boldsymbol{\bf{h}}_{k'l'}|}{\sqrt{\|\boldsymbol{\bf{h}}_{kl}\|^2~ \|\boldsymbol{\bf{h}}_{k'l'}\|^2}}= \frac{|\boldsymbol{\bf{h}}_{kl}^H \boldsymbol{\bf{h}}_{k'l'}|}{\|\boldsymbol{\bf{h}}_{kl}\|~ \|\boldsymbol{\bf{h}}_{k'l'}\|} \to 0. 
\end{split}
\end{equation}
Based on \eqref{8} and \eqref{9}, and by dividing both sides of \eqref{7} with $\xi_k$, the resulting signal for UE $k$ reduces to
\begin{equation}
y_k[n]\to \sqrt{p_k\sum_{j=1}^{L_k}\|\boldsymbol{\bf{h}}_{kj}\|^2} s_k[n-n_{k, \text{max}}] +z_k[n], k=1,...,K.  \label{10}            
\end{equation}

%检查括号的序号
It follows  from \eqref{10} that for multi-user single-carrier DAM with the simple delay pre-compensation and per-path-based MRT beamforming,  the resulting signal of UE $k$ only includes the desired  symbol sequence $s_k[n]$ with one single delay  $n_{k,\text{max}}$, while still achieving the channel gain contributed  by all the  $L_k$ paths. As a result, the original time-dispersive multi-user channels with both ISI and IUI  have been transformed to $K$ parallel  ISI-free and IUI-free AWGN channels, without relying on  the conventional techniques like  channel equalization or multi-carrier transmission. Moreover, based on the received signal in \eqref{10}, the resulting signal-to-noise
ratio (SNR) at UE $k$ is given by $\gamma_{k}^{\text{asym}} =p_k{\|\bar{\boldsymbol{\bf{h}}}_{k}\|^2}/{\sigma^2}$, where $\bar{\boldsymbol{\bf{h}}}_{k} = [\boldsymbol{\bf{h}}_{k1}^H,...,\boldsymbol{\bf{h}}_{kL_k}^H]^H \in \mathbb{C}^{M_tL_k\times1}$. In this case, the optimal power allocation across UEs to maximize the asymptotic sum rate of all the $K$ UEs is readily  obtained by the classical water-filling (WF) strategy \cite{goldsmith}.

%{\color{red}Add a paragraph to give an overivew of what to do next. Start with somthing like:}

In the following, we investigate the performance of multi-user DAM when the BS is equipped with a finite number of antennas. The optimal rate region and the achievable sum rate with classical beamforming strategies are analyzed and compared with benchmarking schemes. 
%新的开始
\section{Rate Region  for Multi-User DAM with Optimal Beamforming}\label{sec5}

\subsection{Achievable Rate Region with Optimal  Beamforming} \label{RRDAM}

To characterize the achievable  rate region for the multi-user DAM system, we first  derive the signal-to-interference-plus-noise ratio (SINR) for the  received signal in \eqref{4}, by treating the ISI and IUI as noise. Since the ISI and IUI in \eqref{4} may have correlated terms, the symbols need to be properly grouped by considering the delay differences of the signal components \cite{DAM}. To this end, let $\Delta_{kl,k'l'}=n_{kl}-n_{k'l'}$  denote the {\it delay difference} between multi-path $l$ of UE $k$ and multi-path $l'$ of UE $k'$. For a given UE pair $(k,k')$, we have $ \Delta_{kl,k'l'}\in \{\Delta_{kk',\min},\Delta_{kk',\min}+1,...,\Delta_{kk',\max} \}$, where $\Delta_{kk',\min} =n_{k, \text{min}}- n_{k', \text{max}}$ and $\Delta_{kk',\max} =n_{k,\text{max}}- n_{k', \text{min}}$. Then for each UE pair $(k,k')$ and delay difference $i\in  \{\Delta_{kk',\min},\Delta_{kk',\min}+1,...,\Delta_{kk',\max} \}$, we define the following effective channel
 \begin{equation}
\boldsymbol{\bf{g}}_{kk'l'} [i] = \label{RRDAM-1}
\begin{cases}
\boldsymbol{\bf{h}}_{kl}, &\text{if}~\exists l\in \{1,...,L_k\}, ~\mbox{s.t.}~\Delta_{kl,k'l'}=i, \\ 
\boldsymbol{\bf{0}} , &\mbox{otherwise}.
\end{cases}
\end{equation}

Then, the received signal in \eqref{4} is equivalently rewritten as % 用户不同，但是可能同时到达接收端，这个上行单用户MIMO也是不同的，这里不可以有不同的延时
\begin{align}
&y_k[n]={\left(\begin{matrix}\sum_{l=1}^{L_k}\end{matrix}\boldsymbol{\bf{h}}_{kl}^H \boldsymbol{\bf{f}}_{kl}\right) s_k[n-n_{k, \text{max}}]}  \label{RRDAM-2} \\  \notag
                 & +{ \sum_{i=\Delta_{kk,\min},i\ne 0}^{\Delta_{kk,\max}}  \left(\sum_{l'=1}^{L_k}\boldsymbol{\bf{g}}_{kkl'}^H[i] \boldsymbol{\bf{f}}_{kl'}\right) s_k[n-n_{k, \text{max}}-i]} \\  \notag
                 &+{\sum_{k'\ne k}^{K}\sum_{i=\Delta_{kk',\min}}^{\Delta_{kk',\max}}\left(\sum_{l'=1}^{L_{k'}} \boldsymbol{\bf{g}}_{kk'l'}^H[i] \boldsymbol{\bf{f}}_{k'l'}\right)  s_{k'}[n-n_{k', \text{max}}-i] }\\  \notag
                 & +z_k[n].
\end{align}

\newcounter{TempEqCnt1} % 创建临时变量TempEqCnt
\setcounter{TempEqCnt1}{\value{equation}} % 将当前公式序号 赋给TempEqCnt
\setcounter{equation}{12} % 当前公式序号变为x，x等于长公式应有的序号减1.
\begin{figure*}[ht] %hb代表放在文章底部，%ht为放在文章顶部
\begin{equation}
\begin{split}
\gamma_k&= \frac{|\sum_{l=1}^{L_k}\boldsymbol{\bf{h}}_{kl}^H \boldsymbol{\bf{f}}_{kl}|^2}{ \sum_{i=\Delta_{kk,\min},i\ne 0}^{\Delta_{kk,\max}}  |\sum_{l'= 1}^{L_k} \boldsymbol{\bf{g}}_{kkl'}^H[i]\boldsymbol{\bf{f}}_{kl'} |^2+ \sum_{k'\ne k}^K\sum_{i=\Delta_{kk',\min}}^{\Delta_{kk',\max}}   |\sum_{l'=1}^{L_{k'}}\boldsymbol{\bf{g}}_{kk'l'}^H[i] \boldsymbol{\bf{f}}_{k'l'}|^2+\sigma^2 } \label{RRDAM-3}\\ 
&=\frac{|\bar{\boldsymbol{\bf{h}}}_k^H\bar{\boldsymbol{\bf{f}}}_k|^2}{\sum_{i=\Delta_{kk,\min},i\ne 0}^{\Delta_{kk,\max}} |\bar{\boldsymbol{\bf{g}}}_{kk}^H[i]\bar{\boldsymbol{\bf{f}}}_k|^2 +\sum_{k'\ne k}^{K} \sum_{i=\Delta_{kk',\min}}^{\Delta_{kk',\max}}  |\bar{\boldsymbol{\bf{g}}}_{kk'}^H[i]\bar{\boldsymbol{\bf{f}}}_{k'}|^2 + \sigma^2 }\\
&=\frac{|\bar{\boldsymbol{\bf{h}}}_k^H\bar{\boldsymbol{\bf{f}}}_k|^2}{ ||\boldsymbol{\bf{G}}_{kk}^H\bar{\boldsymbol{\bf{f}}}_k||^2 +\sum_{k'\ne k}^{K} ||\boldsymbol{\bf{G}}_{kk'}^H\bar{\boldsymbol{\bf{f}}}_{k'}||^2+\sigma^2 }.
\end{split}
\end{equation}
\hrulefill
\end{figure*}
%\vspace{-0.4cm}
The  resulting SINR is given  in~\eqref{RRDAM-3}, shown at the top of the next page, where we have defined:
\begin{align}
\bar{\boldsymbol{\bf{f}}}_{k}&\triangleq [\boldsymbol{\bf{f}}_{k1}^H,...,\boldsymbol{\bf{f}}_{kL_{k}}^H]^H\in \mathbb{C}^{M_tL_{k}\times1}, \label{RRDAM-4} \\
\bar{\boldsymbol{\bf{g}}}_{kk'}[i]&\triangleq \Big[\boldsymbol{\bf{g}}_{kk'1}^H[i],...,\boldsymbol{\bf{g}}_{kk'L_{k'}}^H[i]\Big]^H\in \mathbb{C}^{M_tL_{k'} \times 1 },  \\
 \boldsymbol{\bf{G}}_{kk'} &\triangleq \Big[\bar{\boldsymbol{\bf{g}}}_{kk'}[\Delta_{kk',\min}],...,\bar{\boldsymbol{\bf{g}}}_{kk'}[\Delta_{kk',\max}]\Big], 
\end{align}
with  $  \boldsymbol{\bf{G}}_{kk'}\in \mathbb{C}^{M_tL_{k'} \times {\Delta_{kk', \text{span}}} }$, and $\Delta_{kk', \text{span}}=\Delta_{kk',\max} -\Delta_{kk',\min}$.

Therefore, with the per-path-based transmit beamforming vectors $(\bar{\boldsymbol{\bf{f}}}_1,...,\bar{\boldsymbol{\bf{f}}}_K)$, the achievable rate for UE $k$ is $R_k(\bar{\boldsymbol{\bf{f}}}_1,...,\bar{\boldsymbol{\bf{f}}}_K)=\text{log}_2(1+ \gamma_k)$, and the achievable rate region is the union  of all rate-tuples  that is  achieved by the $K$ users, given by
\begin{align}
\mathcal{R}\triangleq\bigcup_{\sum_{k=1}^K ||\bar{\boldsymbol{\bf{f}}}_{k}||^2 \leq P}\left\{(r_1,...,r_K): r_k= R_k( \bar{\boldsymbol{\bf{f}}}_{1},...,\bar{\boldsymbol{\bf{f}}}_{K}), \forall k\right\}. \label{RRDAM-7}
\end{align}
The outermost boundary of $\mathcal{R}$ is known as  the Pareto boundary \cite{pareto1}.
Similar  to \cite{pareto1} and \cite{pareto2}, the Pareto boundary of the rate region $\mathcal{R}$ can be obtained via solving the following optimization problem for a given set of rate-profile vectors, $\boldsymbol{\alpha}=(\alpha_1,..,\alpha_K)$:
\begin{align} %\tag{\ref{RRDAM-8}{a}} \label{RRDAM-8a}
\max\limits_{R, \{\bar{\boldsymbol{\bf{f}}}_k\}_{k=1}^K} ~&R \label{RRDAM-8}\\ \notag 
\text{ s.t.} ~&\text{log}_2(1+ \gamma_k)\geq\alpha_k R,~ k=1,...,K, \\  \notag 
&\sum_{k= 1 }^K||\bar{\boldsymbol{\bf{f}}}_k||^2\leq P,
\end{align}
where $\alpha_k \geq 0, \forall k$ denotes the weighting coefficient for UE $k$, with $\sum_{k=1}^K\alpha_k=1$. For any given $\boldsymbol{\alpha}$, let $R^{\star}$ denote optimal value to problem \eqref{RRDAM-8}.  Then the rate-tuple $R^{\star}\boldsymbol{\alpha}$ is the intersection between the Pareto boundary of the rate region $\mathcal{R}$ and a ray in the direction of $\boldsymbol{\alpha}$. After solving
problem \eqref{RRDAM-8} with different rate-profile vectors $\boldsymbol{\alpha}$, the complete Pareto boundary of $\mathcal{R}$ can be found.
 %****************************************************************************************************************************************

To solve problem \eqref{RRDAM-8} with any given rate-profile vector $\boldsymbol{\alpha}$ and target rate $R$, it can be transformed into the feasibility problem, given by
\begin{align}
 \text{Find} ~&\bar{\boldsymbol{\bf{f}}}_1,...,\bar{\boldsymbol{\bf{f}}}_K \label{gg2} \\ \notag 
\text{ s.t.}~&\text{log}_2(1+ \gamma_k)\geq\alpha_k R,~ k=1,...,K,\\ \notag
& \sum_{k= 1 }^K||\bar{\boldsymbol{\bf{f}}}_k||^2\leq P.
\end{align}
Let $\gamma_k^{D}=2^{\alpha_k R}-1$, we consider the following problem to minimize the BS transmit power, subject to the minimum SINR constraint for each UE, which is equivalent to problem \eqref{gg2}, i.e.,
\begin{align}
 \min_{\{\bar{\boldsymbol{\bf{f}}}_k\}_{k=1}^K} ~&\sum_{k= 1 }^K||\bar{\boldsymbol{\bf{f}}}_k||^2 \label{RRDAM-10}\\ \notag 
\text{ s.t.}~&\Im{(\bar{\boldsymbol{\bf{h}}}_{k}^H\bar{\boldsymbol{\bf{f}}}_{k})} = 0,~ k=1,...,K,\\ \notag
& \sqrt{\gamma_k^D}
\begin{Vmatrix}
\boldsymbol{\bf{G}}_{kk}^H\bar{\boldsymbol{\bf{f}}}_k \\
\boldsymbol{\bf{G}}_{k1}^H\bar{\boldsymbol{\bf{f}}}_{1} \\
...\\
\boldsymbol{\bf{G}}_{k(k-1)}^H \bar{\boldsymbol{\bf{f}}}_{k-1}\\
\boldsymbol{\bf{G}}_{k(k+1)}^H \bar{\boldsymbol{\bf{f}}}_{k+1} \\
...\\
\boldsymbol{\bf{G}}_{kK}^H \bar{\boldsymbol{\bf{f}}}_{K}\\
\sigma
\end{Vmatrix}
\leq \Re{(\bar{\boldsymbol{\bf{h}}}_{k}^H\bar{\boldsymbol{\bf{f}}}_{k})}, k=1,...,K,
\end{align}
where without loss of optimality, we set $\bar{\boldsymbol{\bf{h}}}_{k}^H\bar{\boldsymbol{\bf{f}}}_{k}$ as a real number,  since any common phase rotation to all elements in $\bar{\boldsymbol{\bf{f}}}_{k}$ does not alter the SINR $\gamma_k$ in \eqref{RRDAM-3} \cite{beam}.
It is observed that problem \eqref{RRDAM-10} is a convex SOCP problem, which can be  efficiently  solved by standard convex optimization  tools such as CVX \cite{CVX}. If the obtained BS transmit  power in problem \eqref{RRDAM-10} is no greater than $P$, then the optimal objective value of problem \eqref{RRDAM-8} satisfies $R^{\star}\geq R$; otherwise, $R^{\star}< R$. Thus, the original problem \eqref{RRDAM-8} can be solved by efficient bisection search over $R$, together with the SOCP problem \eqref{RRDAM-10} with given $R$. The key steps are summarized in Algorithm \ref{255}. Note that in Step 1, the upper bound $R_U$ is obtained by ignoring the ISI and IUI in \eqref{RRDAM-3}, for which the beamforming vectors $\bar{\boldsymbol{\bf{f}}}_{k}$ simply matches with the channel $\bar{\boldsymbol{\bf{h}}}_{k}$.

In the following, for performance comparison, we study the achievable rate regions of two benchmarking schemes, namely the  alternative single-carrier scheme termed {\it strongest-path-based (SP) beamforming}\cite{TDL}, which only exploits the single strongest path of each UE as the desired signal, and the multi-carrier OFDM scheme. 

\subsection{Achievable Rate Region by the Strongest-Path-Based Beamforming} \label{RR1}

With the strongest-path-based beamforming, only a single beam is used for each user. Specifically, let $\boldsymbol{\bf{f}}_k \in \mathbb{C}^{M_t \times 1}$ denote the transmit beamforming vector of UE $k$. Then the transmitted signal 
\begin{equation}
\boldsymbol{\bf{x}}_{\text{SP}}[n]=\sum_{k=1}^{K}\boldsymbol{\bf{f}}_k s_k[n] \label{RRSP-1}.
\end{equation} 
%By comparing  \eqref{2} and \eqref{RRSP-1}, it is observed  that different from DAM, the strongest path-beamforming scheme uses only one single beam for each user. 

Without loss of generality, we assume that the first path of each UE is the strongest path. Thus, with the channel model in \eqref{1}, the received signal at UE $k$ for the strongest-path-based beamforming  is expressed as 
\begin{align}
y_{k,\text{SP}}[n]&=\boldsymbol{\bf{h}}_k^H[n]*\boldsymbol{\bf{x}}_{\text{SP}}[n]+{z}_k[n] \label{RRSP-2}\\   \notag  %~\eqref{111}
&=\underbrace{ \boldsymbol{\bf{h}}_{k1}^H \boldsymbol{\bf{f}}_k s_k[n-n_{k1}]}_{\text{Desired signal}} + \underbrace{ \sum_{ l\ne 1}^{L_k} \boldsymbol{\bf{h}}_{kl}^H  \boldsymbol{\bf{f}}_k s_{k}[n-n_{kl}]}_{\text{ISI}} \\  \notag
                  & +\underbrace{ \sum_{l=1}^{L_k}\sum_{k'\ne k }^K  \boldsymbol{\bf{h}}_{kl}^H \boldsymbol{\bf{f}}_{k'} s_{k'}[n-n_{kl}]}_{\text{IUI}}+ z_k[n].
\end{align}
In this case, only the single strongest path signal is regarded as the desired signal, and all the remaining $L_{\mathrm{tot}}-1$  multi-paths cause the detrimental ISI or IUI. The resulting SINR at UE $k$ based on \eqref{RRSP-2} is 
\begin{equation}
\gamma_{k}^{\text{SP}}=\frac{|\boldsymbol{\bf{h}}_{k1}^H \boldsymbol{\bf{f}}_k |^2}{ \sum_{l\ne 1}^{L_k} |\boldsymbol{\bf{h}}_{kl}^H  \boldsymbol{\bf{f}}_k|^2+ \sum_{l=1}^{L_{k}}  \sum_{k'\ne k}^K  |\boldsymbol{\bf{h}}_{kl}^H \boldsymbol{\bf{f}}_{k'}|^2+\sigma^2  }. \label{RRSP-3}
\end{equation}

Note that when $M_t \gg L_{\mathrm{tot}}$, by applying MRT beamforming,  with $\boldsymbol{\bf{f}}_{k}=\sqrt{\bar{p}_k}\frac{\boldsymbol{\bf{h}}_{k1}}{\|\boldsymbol{\bf{h}}_{k1}\|}$, where $\bar{p}_k$ is the power allocation of  UE $k$,  the received signal in \eqref{RRSP-2} reduces to
\begin{equation}
y_{k,\text{SP}}[n] \to \sqrt{\bar{p}_k}\|\boldsymbol{\bf{h}}_{k1}\| s_k[n-n_{k1}] + z_k[n].   \label{dam22}              
\end{equation}
As compared with DAM, both ISI and IUI approach to zero asymptotically for the strongest-path-based beamforming, but the power for UE $k$ is only contributed by one single path, rather than all the $L_k$ multi-path as in DAM. The achievable rate region for the  strongest-path-based beamforming can be similarly  obtained with Algorithm \ref{255} by replacing the SINR expression  \eqref{RRDAM-3} as \eqref{RRSP-3}.

 \begin{algorithm}[t]
  \caption{Solving Problem \eqref{RRDAM-8} to Find A Pareto-Optimal Point for Multi-User DAM  Rate Region $\mathcal{R}$}
  \label{255}
  \begin{algorithmic}[1]
   \State {\textbf{Initialize}: Set the lower bound $R_L = 0$ and the upper  bound $R_U = \min\limits_{k=1,...,K}\{ R_k^{\star} /\alpha_k\}$, where $R_k^{\star}= \text{log}_2(1+P||\bar{\boldsymbol{\bf{h}}}_{k}||^2/\sigma^2)$}, and a small threshold $\epsilon$ and rate-profile vectors $\boldsymbol{\alpha}$.  
   \While {$(R_U-R_L)>\epsilon R_L$}
   \State Set $R = (R_L + R_U)/2$;
   \State Solve the SOCP problem \eqref{RRDAM-10}, and denote the optimal 
   \Statex\hspace{4 mm} objective value as $P^{\star}$ and the resulting  optimal 
   \Statex\hspace{4 mm} solution as $\{\bar{\boldsymbol{\bf{f}}}_k\}_{k=1}^K $. If $P^{\star}\leq P$, then let $R_L=R$, 
   \Statex\hspace{4 mm} and $\bar{\boldsymbol{\bf{f}}}_k^{\star}=\bar{\boldsymbol{\bf{f}}}_k$; otherwise, let $R_U = R$.
    \EndWhile
    \Ensure  
         $R^{\star}=R_L $ and $\{\bar{\boldsymbol{\bf{f}}}_k^{\star}\}_{k=1}^K$.
  \end{algorithmic}
\end{algorithm}

\subsection{Achievable Rate Region by OFDM}\label{B2}

For the benchmarking OFDM scheme, let $M$ denote the number of sub-carriers. With the channel impulse response given in \eqref{1}, the channel of the $m$th sub-carrier for UE $k$ can be obtained by applying $M$-point DFT, given by
\begin{equation}
\boldsymbol{\bf{h}}_{k,m}=\frac{1}{\sqrt{M}}\sum_{l=1}^{L_k}\boldsymbol{\bf{h}}_{kl}e^{-j\frac{2\pi mn_{kl}}{M}} \label{RROFDM-1}.
\end{equation}

The transmitted signal by the BS of the $m$th sub-carrier can be expressed as 
\begin{equation}
\boldsymbol{\bf{x}}_m=\begin{matrix}\sum_{k=1}^{K}\end{matrix}\boldsymbol{\bf{d}}_{k,m} s_{k,m},\label{RROFDM-2}
\end{equation}
where $\boldsymbol{\bf{d}}_{k,m} \in \mathbb{C}^{M_t \times 1}$ and ${{s}}_{k,m}$ denote the frequency-domain beamforming vector and the information-bearing symbol for UE $k$ at the $m$th sub-carrier, respectively, with $\mathbb{E}[|s_{k,m}|^2]=1$. The corresponding  power constraint  is
\begin{equation}
\begin{matrix}\sum_{m=1}^M\end{matrix}\mathbb{E}[\|\boldsymbol{\bf{x}}_m\|^2]=\begin{matrix}\sum_{m=1}^M\end{matrix}\begin{matrix}\sum_{k=1}^{K}\end{matrix}\|\boldsymbol{\bf{d}}_{k,m}\|^2\leq MP \label{RROFDM-3}.
\end{equation} 
After removing the cyclic prefix (CP) whose length is no smaller than the maximum delay spread of all UEs, the frequency-domain received  signal  of UE $k$ at the $m$th sub-carrier is
\begin{align}
y_{k,m}&=\boldsymbol{\bf{h}}_{k,m}^H\sum_{k'=1}^{K}\boldsymbol{\bf{d}}_{k',m} s_{k',m} + {z}_{k,m} \label{RROFDM-4}\\   \notag
&=\underbrace{\boldsymbol{\bf{h}}_{k,m}^H\boldsymbol{\bf{d}}_{k,m} s_{k,m}}_{\text{Desired signal}} + \underbrace{\boldsymbol{\bf{h}}_{k,m}^H\begin{matrix}\sum_{k'\ne k}^{K}\end{matrix}\boldsymbol{\bf{d}}_{k',m} s_{k',m}}_{\text{IUI}} + {z}_{k,m},\\   \notag
\end{align}
where ${z}_{k,m}\sim \mathcal{CN}(0,\hat{\sigma}^2)$  is the AWGN with $\hat{\sigma}^2={\sigma^2}/{M}$.
The corresponding SINR is expressed as
\begin{equation}
\gamma_{k,m}=\frac{|\boldsymbol{\bf{h}}_{k,m}^H\boldsymbol{\bf{d}}_{k,m}|^2}{\sum_{k'\ne k}^K |\boldsymbol{\bf{h}}_{k,m}^H\boldsymbol{\bf{d}}_{k',m}|^2+\hat{\sigma}^2}.\label{RROFDM-5}
\end{equation}
The  achievable rate for UE $k$  without considering the CP overhead is thus $R_k=\frac{1}{M}\sum_{m=1}^M\text{log}_2(1+\gamma_{k,m})$.
Similar to Section \ref{RRDAM}, by using  different rate-tuples $\boldsymbol{\bf{\alpha}} = (\alpha_{1},..., \alpha_{K})$, the Pareto boundary  of  the achievable rate region for OFDM can be obtained by solving the following optimization problem
\begin{align}
&\max\limits_{{\mu, \boldsymbol{\bf{d}}}_{k,m}, \forall k,m} \mu\label{RROFDM-7}\\ \notag 
\text{ s.t.} ~&\sum_{m=1}^M\text{log}_2\bigg(1+ \frac{|\boldsymbol{\bf{h}}_{k,m}^H\boldsymbol{\bf{d}}_{k,m}|^2}{\sum_{k'\ne k}^K |\boldsymbol{\bf{h}}_{k,m}^H\boldsymbol{\bf{d}}_{k',m}|^2+\hat{\sigma}^2}\bigg)\geq M \alpha_{k} \mu,\forall k,\\ \notag
&\begin{matrix}\sum_{m=1}^M\end{matrix} \begin{matrix}\sum_{k= 1 }^K\end{matrix}||{\boldsymbol{\bf{d}}}_{k,m}||^2\leq MP.
\end{align}

Problem \eqref{RROFDM-7} is non-convex since the first constraint is non-convex. Fortunately,  successive convex approximation (SCA) technique can be  applied to obtain an efficient approximate solution \cite{SCA}. To this end, problem \eqref{RROFDM-7} is equivalently written as  
\begin{align}
&\max\limits_{{\mu, \boldsymbol{\bf{d}}}_{k,m}, \forall k,m} \mu \label{RROFDM-8} \\ \notag 
\text{ s.t.}~&\sum_{m=1}^M\text{log}_2\bigg(\sum_{k'=1}^K |\boldsymbol{\bf{h}}_{k,m}^H\boldsymbol{\bf{d}}_{k',m}|^2+\hat{\sigma}^2\bigg)-\\ \notag 
&\sum_{m=1}^M\text{log}_2\bigg(\sum_{k'\ne k}^K |\boldsymbol{\bf{h}}_{k,m}^H\boldsymbol{\bf{d}}_{k',m}|^2+\hat{\sigma}^2\bigg)\geq M\alpha_{k}\mu,~ \forall k,m,\\ \notag
& \begin{matrix}\sum_{m=1}^M\end{matrix} \begin{matrix}\sum_{k= 1 }^K\end{matrix}||{\boldsymbol{\bf{d}}}_{k,m}||^2\leq MP.
\end{align}

By introducing slack variables $S_{kk',m}$ and $C_{kk',m}$,  problem \eqref{RROFDM-8} can be equivalently  reformulated as
\begin{align}
&\max\limits_{{\mu, \boldsymbol{\bf{d}}}_{k,m}, S_{kk',m}, C_{kk',m}, \forall k,k',m} \mu \label{RROFDM-9} \\ \notag 
\text{ s.t.}~&\sum_{m=1}^M\text{log}_2\bigg(\sum_{k'=1}^K S_{kk',m}+\hat{\sigma}^2\bigg)-\\ \notag 
&\sum_{m=1}^M\text{log}_2\bigg(\sum_{k'\ne k}^KC_{kk',m} +\hat{\sigma}^2\bigg)\geq M\alpha_{k}\mu,~ \forall k,m,\\ \notag
&S_{kk',m}\leq|\boldsymbol{\bf{h}}_{k,m}^H\boldsymbol{\bf{d}}_{k',m}|^2,\forall k,k',m,\\ \notag
& C_{kk',m} \geq|\boldsymbol{\bf{h}}_{k,m}^H\boldsymbol{\bf{d}}_{k',m}|^2, \forall k,k',m,\\ \notag
& \begin{matrix}\sum_{m=1}^M\end{matrix} \begin{matrix}\sum_{k= 1 }^K\end{matrix}||{\boldsymbol{\bf{d}}}_{k,m}||^2\leq MP.
\end{align}
Note that problem \eqref{RROFDM-9} is equivalent to problem \eqref{RROFDM-8}, since there exists an optimal solution to problem \eqref{RROFDM-9} such that the second and third contraints are met with equality. Otherwise, we can always increase $S_{kk',m}$ or decrease $C_{kk',m}$ without decreasing the objective value $\mu$ nor violating the constraints. Since $|\boldsymbol{\bf{h}}_{k,m}^H\boldsymbol{\bf{d}}_{k',m}|^2$ is  a convex differentiable function with respect to $\boldsymbol{\bf{d}}_{k',m}$, it is
 globally lower bounded by its first-order Taylor expansion, given by 
\begin{align}
|\boldsymbol{\bf{h}}_{k,m}^H\boldsymbol{\bf{d}}_{k',m}|^2\geq  &|\boldsymbol{\bf{h}}_{k,m}^H\boldsymbol{\bf{d}}_{k',m}^{(r)}|^2+2\Re\Big\{(\boldsymbol{\bf{d}}_{k',m}^{(r)})^H\boldsymbol{\bf{h}}_{k,m}\boldsymbol{\bf{h}}_{k,m}^H\label{RROFDM-10}\notag \\  
&\times (\boldsymbol{\bf{d}}_{k',m}-\boldsymbol{\bf{d}}_{k',m}^{(r)})\Big\},
\end{align}
where $\boldsymbol{\bf{d}}_{k',m}^{(r)}$ is the obtained  beamforming vector at the $r$-th iteration.
Similarly, $\text{log}_2(\sum_{k'\ne k}^KC_{kk',m} +\hat{\sigma}^2)$ is a concave  function   with respect to $C_{kk',m}$, which is globally upper bounded by its first-order Taylor expansion, given by 
\begin{align}
&\text{log}_2\left(\begin{matrix}\sum_{k'\ne k}^K\end{matrix}C_{kk',m} +\hat{\sigma}^2\right) \leq \text{log}_2\left(\begin{matrix}\sum_{k'\ne k}^K\end{matrix}C_{kk',m}^{(r)} +\hat{\sigma}^2\right) \label{RROFDM-11} \notag \\  
&+\sum_{k'\ne k}^K\frac{\text{log}_2e}{\sum_{i\ne k}^KC_{ki,m}^{(r)} +\hat{\sigma}^2}(C_{kk',m}-C_{kk',m}^{(r)}),
\end{align}
where $C_{kk',m}^{(r)}$ is the  given local point at the $r$-th iteration.

By replacing the global lower and upper bounds in \eqref{RROFDM-10} and \eqref{RROFDM-11}, problem \eqref{RROFDM-9} can be transformed to  problem \eqref{RROFDM-12}, shown at the top of this page.
Since problem \eqref{RROFDM-12} is convex, it can be efficiently solved by  convex optimization toolbox such as CVX. Moreover, the efficient local solution to problem \eqref{RROFDM-9} can be obtained by successively updating the local point $\{\boldsymbol{\bf{d}}_{k',m}^{(r)}, C_{kk',m}^{(r)}, S_{kk',m}^{(r)}, \forall k,k',m\}$. The details for solving problem \eqref{RROFDM-9}  are summarized in Algorithm \ref{scapareto}. Since the  objective value of problem \eqref{RROFDM-12} is non-decreasing over each iteration, Algorithm \ref{scapareto} is guaranteed to converge \cite{SCA}.

\newcounter{TempEqCnt2} % 创建临时变量TempEqCnt
\setcounter{TempEqCnt2}{\value{equation}} % 将当前公式序号 赋给TempEqCnt
\setcounter{equation}{34} % 当前公式序号变为x，x等于长公式应有的序号减1.
\begin{figure*}[ht] %hb代表放在文章底部，%ht为放在文章顶部
\begin{equation}
\begin{split}
&\max\limits_{{\mu, \boldsymbol{\bf{d}}}_{k,m}, S_{kk',m}, C_{kk',m}, \forall k,k',m} \mu \label{RROFDM-12} \\ 
\text{ s.t.}~&\sum_{m=1}^M\left\{\text{log}_2\left(\sum_{k'=1}^K S_{kk',m}+\hat{\sigma}^2\right)-\text{log}_2\left(\sum_{k'\ne k}^KC_{kk',m}^{(r)} +\hat{\sigma}^2\right)-\sum_{k'\ne k}^K\frac{\text{log}_2e}{\sum_{i\ne k}^KC_{ki,m}^{(r)} +\hat{\sigma}^2}(C_{kk',m}-C_{kk',m}^{(r)})\right\}\geq M\alpha_{k}\mu,~ \forall k,\\ 
  & S_{kk',m}\leq |\boldsymbol{\bf{h}}_{k,m}^H\boldsymbol{\bf{d}}_{k',m}^{(r)}|^2+2\Re\left\{(\boldsymbol{\bf{d}}_{k',m}^{(r)})^H\boldsymbol{\bf{h}}_{k,m}\boldsymbol{\bf{h}}_{k,m}^H(\boldsymbol{\bf{d}}_{k',m}-\boldsymbol{\bf{d}}_{k',m}^{(r)})\right\}, \forall k,k',m, \\ 
& C_{kk',m} \geq|\boldsymbol{\bf{h}}_{k,m}^H\boldsymbol{\bf{d}}_{k',m}|^2, \forall k,k',m,\\ 
& \begin{matrix}\sum_{m=1}^M\end{matrix} \begin{matrix}\sum_{k= 1 }^K\end{matrix}||{\boldsymbol{\bf{d}}}_{k,m}||^2\leq MP.
\end{split}
\end{equation}
\hrulefill
\end{figure*}
\begin{algorithm}[t]
  \caption{SCA-based Algorithm for Problem \eqref{RROFDM-9}}
  \label{scapareto}
  \begin{algorithmic}[1]
   \State Initialize a feasible solution $\boldsymbol{\bf{d}}_{k',m}^{(0)}, C_{kk',m}^{(0)}, S_{kk',m}^{(0)},  $
    \Statex $\forall k,k',m$ to problem \eqref{RROFDM-9}. Let $r=0$.
   \Repeat
   \State {Solve the convex optimization  problem \eqref{RROFDM-12} for given }
   \Statex\hspace{5.5mm}{local points $\{\boldsymbol{\bf{d}}_{k',m}^{(r)}, C_{kk',m}^{(r)}, S_{kk',m}^{(r)}\}$, and denote the  
    \Statex\hspace{4mm} optimal solution as $\{\boldsymbol{\bf{d}}_{k',m}^{(r+1)}, C_{kk',m}^{(r+1)}, S_{kk',m}^{(r+1)}\}$. }
    \State Update $r=r+1$.
    \Until{ The  fractional increase of objective value of problem \eqref{RROFDM-9}  is below a certain threshold}.
  \end{algorithmic}
\end{algorithm}

%yyyyyyyyyyyyyyyyyyyyyyyyyyyyyyyyyyyyyy
\section{ Sum Rate for  Multi-User DAM  and Benchmarking  Schemes} \label{sec2}
%\vspace{-0.1cm}
In this section, we study the  sum rate of multi-user DAM system by considering  the three classical low-complexity  beamforming  schemes in a per-path basis, namely per-path-based MRT, ZF, and RZF, respectively. The corresponding  benchmarking schemes of the strongest-path-based beamforming and OFDM are also considered for performance comparison.

%\vspace{-0.4cm}
\subsection{Sum Rate of Multi-User DAM}
\subsubsection{Per-Path-Based MRT Beamforming} 

%\vspace{-0.4cm}
The per-path-based MRT beamforming for the  asymptotic analysis with $M_t\gg L_{\mathrm{tot}}$ has been considered  in Section \ref{sec1}. For the general case  with finite $M_t$, the low complexity per-path-based MRT beamforming for multi-path $l$ of UE $k$ is given by
\begin{equation}
\boldsymbol{\bf{f}}_{kl}^{\text{MRT}}=\sqrt{P}{\boldsymbol{\bf{h}}_{kl}}/{\|\boldsymbol{\bf{H}}\|_{F}},\label{SRDAMMRT-1}
\end{equation}
where $\boldsymbol{\bf{H}}=[\boldsymbol{\bf{h}}_{11,}...,\boldsymbol{\bf{h}}_{1L_1},...,\boldsymbol{\bf{h}}_{K1,}...,\boldsymbol{\bf{h}}_{KL_K}] \in \mathbb{C}^{M_t \times L_{\mathrm{tot}}}$. Let $\bar{\boldsymbol{\bf{f}}}_{k}^{\text{MRT}}=\big[(\boldsymbol{\bf{f}}_{k1}^{\text{MRT}})^H,...,(\boldsymbol{\bf{f}}_{kL_k}^{\text{MRT}})^H\big]^H=\sqrt{P}\frac{\bar{\boldsymbol{\bf{h}}}_k}{\|\boldsymbol{\bf{H}}\|_{F}}$, with $\bar{\boldsymbol{\bf{h}}}_k$ defined below \eqref{10}. By  substituting $\bar{\boldsymbol{\bf{f}}}_{k}^{\text{MRT}}$ into \eqref{RRDAM-3}, the resulting SINR is expressed as
\begin{equation}
\gamma_{k}^{\text{MRT}}=\frac{\|\bar{\boldsymbol{\bf{h}}}_k\|^4}{ ||\boldsymbol{\bf{G}}_{kk}^H\bar{\boldsymbol{\bf{h}}}_k||^2 +\sum_{k'\ne k}^{K} ||\boldsymbol{\bf{G}}_{kk'}^H\bar{\boldsymbol{\bf{h}}}_{k'}||^2+\|\boldsymbol{\bf{H}}\|_{F}^2\sigma^2/P }\label{SRDAMMRT-2}.
\end{equation}

The sum rate of the per-path-based MRT beamforming is  thus $\sum_{k=1}^K\text{log}_2(1+\gamma_{k}^{\text{MRT}})$.

%\begin{matrix}\sum_{m=1}^M\end{matrix} \begin{matrix}\sum_{k= 1 }^K\end{matrix}
\subsubsection{Per-Path-Based ZF Beamforming} \label{ZF}

The per-path-based ZF beamforming  $\boldsymbol{\bf{f}}_{kl'}, k = 1,...,K, l'=1,...,L_k,  $ are designed so that the ISI and IUI in \eqref{4} are perfectly eliminated, i.e., 
\begin{align}
\boldsymbol{\bf{h}}_{kl}^H \boldsymbol{\bf{f}}_{kl'}^{\text{ZF}}&=0, ~\forall l'\neq l, \forall k,  \label{SRDAMZF-1}  \\
\boldsymbol{\bf{h}}_{kl}^H \boldsymbol{\bf{f}}_{k'l'}^{\text{ZF}}&=0, ~\forall k'\neq k, \text{and} ~\forall l, l'. \label{SRDAMZF-2}
\end{align}
Denote by $\boldsymbol{\bf{H}}_{kl'}\in \mathbb{C}^{M_t \times (L_{\mathrm{tot}}-1)}$  the submatrix of $\boldsymbol{\bf{H}}$   excluding the column $\boldsymbol{\bf{h}}_{kl'}$. The per-path-based ZF constraints in \eqref{SRDAMZF-1} and \eqref{SRDAMZF-2} can be compactly written as
\begin{equation}
\boldsymbol{\bf{H}}_{kl'}^H \boldsymbol{\bf{f}}_{kl'}^{\text{ZF}}=\boldsymbol{\bf{0}}_{(L_{\mathrm{tot}}-1)\times 1}, ~\forall k, l'. \\ \label{SRDAMZF-3}
\end{equation}
The above ZF constraint is feasible when $M_t\geq L_{\mathrm{tot}}$.

Let $\boldsymbol{\bf{F}}^{\text{ZF}} =[\boldsymbol{\bf{f}}_{11}^{\text{ZF}},...,\boldsymbol{\bf{f}}_{1L_1}^{\text{ZF}},...,\boldsymbol{\bf{f}}_{K1}^{\text{ZF}},...,\boldsymbol{\bf{f}}_{KL_K}^{\text{ZF}}]\in \mathbb{C}^{M_t \times L_{\mathrm{tot}}}$ denote the matrix composed by all the per-path-based ZF beamforming vectors. Without loss of generality, we can decompose  $\boldsymbol{\bf{F}}^{\text{ZF}}$ as $\boldsymbol{\bf{F}}^{\text{ZF}}=\boldsymbol{\bf{W}}\boldsymbol{\bf{V}}^{\frac{1}{2}}$, where $\boldsymbol{\bf{W}} =[{\boldsymbol{\bf{w}}}_{11},...,{\boldsymbol{\bf{w}}}_{1L_1},...,{\boldsymbol{\bf{w}}}_{K1},...,{\boldsymbol{\bf{w}}}_{KL_K}]\in \mathbb{C}^{M_t \times L_{\mathrm{tot}}}$ is designed to guarantee the ZF constraints in \eqref{SRDAMZF-3}, and $\boldsymbol{\bf{V}}=\text{diag}\{v_{11},...,v_{1L_1},...,v_{K1},...,v_{KL_K}\}\in \mathbb{C}^{L_{\mathrm{tot}} \times L_{\mathrm{tot}}}$ with non-negative real-valued diagonal elements denotes   the power allocation matrix satisfying the power constraint $\sum_{k=1}^K \sum_{l'=1}^{L_{k}}  \|\boldsymbol{\bf{f}}_{kl'}^{\text{ZF}}\|^2 \leq P $. One effective solution for $\boldsymbol{\bf{W}}$ to guarantee the ZF constraints  \eqref{SRDAMZF-3} is by letting $\boldsymbol{\bf{H}}^H\boldsymbol{\bf{W}}=\boldsymbol{\bf{I}}_{L_{\mathrm{tot}}}$. When $M_t\geq L_{\mathrm{tot}}$, the matrix $\boldsymbol{\bf{W}}$ can be directly obtained by taking the right pseudo inverse of $\boldsymbol{\bf{H}}^H$, i.e., $\boldsymbol{\bf{W}}=(\boldsymbol{\bf{H}}^H)^{\dagger} = \boldsymbol{\bf{H}}(\boldsymbol{\bf{H}}^H\boldsymbol{\bf{H}})^{-1}$. As a result, the per-path-based ZF beamforming $\boldsymbol{\bf{f}}_{kl}^{\text{ZF}}$ can be expressed as $\boldsymbol{\bf{f}}_{kl}^{\text{ZF}} =\sqrt{v_{kl}}\boldsymbol{\bf{w}}_{kl}$, where $\boldsymbol{\bf{w}}_{kl}$ is the $\big(\sum_{j=1}^{k-1}L_j+l\big)$th column of $\boldsymbol{\bf{W}}$. By substituting $\boldsymbol{\bf{f}}_{kl}^{\text{ZF}}$ into \eqref{4}, the received signal reduces to 
\begin{align} %\begin{matrix}
y_k[n]&=\begin{matrix}\sum_{l=1}^{L_k}\end{matrix}\sqrt{v_{kl}} \boldsymbol{\bf{h}}_{kl}^H\boldsymbol{\bf{w}}_{kl} s_k[n-n_{k, \text{max}}] + z_k[n] \label{SRDAMZF-4}  \\ \notag
&=\left(\begin{matrix}\sum_{l=1}^{L_k}\end{matrix}\sqrt{v_{kl}}\right) s_k[n-n_{k, \text{max}}] + z_k[n], k=1,...,K,
\end{align}
where  $\boldsymbol{\bf{h}}_{kl}^H\boldsymbol{\bf{w}}_{kl}=1$ is used  based on the property of pseudo inverse. It is observed from \eqref{SRDAMZF-4} that similar to \eqref{10}, the original time-dispersive multi-user broadcast channel with ISI and IUI is also transformed into $K$ parallel  ISI- and IUI-free AWGN channels. The SNR of UE $k$  is $\gamma_{k}^{\text{ZF}}={(\sum_{l=1}^{L_k}\sqrt{v_{kl}})^2 }\big/{ \sigma^2}$, 
and the achievable rate of UE $k$  is  $\text{log}_2(1+\gamma_{k}^{\text{ZF}})$.
The optimal power allocation coefficients $v_{kl}, \forall k,l$, to  maximize the sum rate can be found by solving the following problem
\begin{equation}
\begin{split}
\max\limits_{{v}_{kl}, \forall k,l}~ &\sum_{k=1}^K\text{log}_2\left(1+\frac{\left(\sum_{l=1}^{L_k}\sqrt{v_{kl}}\right)^2 }{ \sigma^2}\right)\label{SRDAMZF-5}\\
\text{s.t.}~&\sum_{k=1}^K \sum_{l=1}^{L_k}v_{kl}\|\boldsymbol{\bf{w}}_{kl}\|^2 \leq P,\\
& {v}_{kl}\geq 0, \forall k,l.
\end{split}
\end{equation}
By defining $\bar{v}_{kl}={v_{kl}}\|\boldsymbol{\bf{w}}_{kl}\|^2$, $\boldsymbol{\bf{t}}_{k} =\left[\sqrt{\bar{v}_{k1}},...,\sqrt{\bar{v}_{kL_k}}\right]^T$, and $\boldsymbol{\bf{q}}_{k} =[{1}/{\|\boldsymbol{\bf{w}}_{k1}\|},...,{1}/{\|\boldsymbol{\bf{w}}_{kL_k}\|}]^T$, problem \eqref{SRDAMZF-5} can be equivalently written as
\begin{equation}
\begin{split}
\max\limits_{\{\boldsymbol{\bf{t}}_k\}_{k=1}^K}~ &\sum_{k=1}^K \text{log}_2\left(1+{(\boldsymbol{\bf{t}}_{k}^T\boldsymbol{\bf{q}}_{k})^2 }/{ \sigma^2}\right)\label{SRDAMZF-6}\\
\text{s.t.}~&\sum_{k=1}^K  \|{\boldsymbol{\bf{t}}_{k}}\|^2 \leq P, \\
& {\boldsymbol{\bf{t}}_{k}} \geq  \boldsymbol{0}_{L_k}, \forall k.
\end{split}
\end{equation}

To derive the optimal solution to problem \eqref{SRDAMZF-6}, the auxiliary variables $\{P_k\}_{k=1}^K$ are introduced, and  problem \eqref{SRDAMZF-6} can be equivalently written as 
\begin{equation}
\begin{split}
\max\limits_{\{\boldsymbol{\bf{t}}_k, P_k\}_{k=1}^K}~& \sum_{k=1}^K \text{log}_2\left(1+{(\boldsymbol{\bf{t}}_{k}^T\boldsymbol{\bf{q}}_{k})^2 }/{ \sigma^2}\right)\label{SRDAMZF-7}\\ 
\text{s.t.} ~&\|{\boldsymbol{\bf{t}}_{k}}\|^2 \leq P_k, \forall k,\\ 
&\sum_{k=1}^K P_k\leq P,\\
&{\boldsymbol{\bf{t}}_{k}} \geq  \boldsymbol{0}_{L_k}, \forall k.
\end{split}
\end{equation}

A closer look at problem \eqref{SRDAMZF-7} shows that for any given feasible  $\{P_k\}_{k=1}^K$, the optimal $\boldsymbol{\bf{t}}_{k}$ can be obtained by applying the Cauchy-Schwarz inequality, given by  $\boldsymbol{\bf{t}}_{k}=\sqrt{P_k}{\boldsymbol{\bf{q}}_{k}}/{\|\boldsymbol{\bf{q}}_{k}\|}, \forall k$. As a result, problem \eqref{SRDAMZF-7} reduces to finding the optimal power allocation ${P_k}$, given by  
\begin{equation}
\begin{split}
\max\limits_{ \{P_k\}_{k=1}^K}~ &\sum_{k=1}^K \text{log}_2\left(1+{P_k\|\boldsymbol{\bf{q}}_{k}\|^2 }/{\sigma^2}\right) \\  \label{SRDAMZF-8}
\text{s.t.} ~&\sum_{k=1}^K  P_k \leq P.
\end{split}
\end{equation}
The optimal solution to problem \eqref{SRDAMZF-8} can be  obtained via the classical WF solution, i.e.,
\begin{equation}
P_k=\left(\lambda-{\sigma^2}/{\|\boldsymbol{\bf{q}}_{k}\|^2}\right)^+ \label{SRDAMZF-9},
\end{equation}
where $\lambda$ is the water level to ensure that the power constraint in problem \eqref{SRDAMZF-8} is satisfied with equality. Thus, the sum rate for per-path-based ZF beamforming is obtained accordingly.

\subsubsection{Per-Path-Based RZF Beamforming} \label{RZF}  %弄完自修
To achieve a balance between mitigating the interference suffered by MRT beamforming and the noise enhancement issue suffered by ZF beamforming, we consider the per-path-based RZF beamforming, where the condition $M_t\geq L_{\mathrm{tot}}$ for the per-path-based ZF beamforming is no longer needed. Let $\tilde{\boldsymbol{\bf{F}}}=\boldsymbol{\bf{H}}(\boldsymbol{\bf{H}}^H\boldsymbol{\bf{H}}+\epsilon_{\text{RZF}} \boldsymbol{I}_{L_{\mathrm{tot}}})^{-1}=[\tilde{\boldsymbol{\bf{f}}}_{11},...,\tilde{\boldsymbol{\bf{f}}}_{1L_1},...,\tilde{\boldsymbol{\bf{f}}}_{K1},...,\tilde{\boldsymbol{\bf{f}}}_{KL_K}]$, where $\epsilon_{\text{RZF}}$ is the regularization parameter given by $\epsilon_{\text{RZF}}={L_{\mathrm{tot}}\sigma^2}/{P}$ \cite{RZF}. Then the per-path-based RZF beamforming is set as
\begin{equation}
 \boldsymbol{\bf{f}}_{kl}^{\text{RZF}}=\sqrt{p_{kl}}e^{j\phi_{kl}}{\tilde{\boldsymbol{\bf{f}}}_{kl}}/{\|\tilde{\boldsymbol{\bf{f}}}_{kl}\|}, \label{SRDAMRZF-1}
\end{equation}
where $p_{kl}$ is the power allocated to path $l$ of UE $k$ and $e^{j\phi_{kl}}$ is the phase rotation.  
%\begin{matrix} \sum_{l=1}^{L_k}\boldsymbol{\bf{h}}_{kl}^H \boldsymbol{\bf{f}}_{kl}^{\text{RZF}}\end{matrix}

With \eqref{SRDAMRZF-1}, the desired signal power  in the numerator of  \eqref{RRDAM-3} is expressed as
\begin{equation}
\begin{split}
\left|\begin{matrix}\sum_{l=1}^{L_k}\end{matrix}\boldsymbol{\bf{h}}_{kl}^H \boldsymbol{\bf{f}}_{kl}^{\text{RZF}}\right|^2=\left|\begin{matrix}\sum_{l=1}^{L_k}\end{matrix} \sqrt{p_{kl}}e^{j\phi_{kl}}\boldsymbol{\bf{h}}_{kl}^H{\tilde{\boldsymbol{\bf{f}}}_{kl}}/{\|\tilde{\boldsymbol{\bf{f}}}_{kl}\|}\right|^2=|\boldsymbol{\bf{a}}_k^H\boldsymbol{\bf{u}}_k|^2, \label{SRDAMRZF-2}
\end{split}
\end{equation}
where we define $\boldsymbol{\bf{a}}_k\triangleq[\sqrt{p_{k1}}e^{j\phi_{k1}},...,\sqrt{p_{kL_k}}e^{j\phi_{kL_k}}]^H\in \mathbb{C}^{L_k \times 1}$, and  $\boldsymbol{\bf{u}}_k=\left[\boldsymbol{\bf{h}}_{k1}^H{\tilde{\boldsymbol{\bf{f}}}_{k1}}/{\|\tilde{\boldsymbol{\bf{f}}}_{k1}\|},...,\boldsymbol{\bf{h}}_{kL_k}^H{\tilde{\boldsymbol{\bf{f}}}_{kL_k}}/{\|\tilde{\boldsymbol{\bf{f}}}_{kL_k}\|}\right]^H \in \mathbb{C}^{L_k \times 1}$. Similarly,  the ISI power in the denominator of \eqref{RRDAM-3} is expressed as
\begin{align}
&\begin{matrix}\sum_{i=\Delta_{kk,\min},i\ne 0}^{\Delta_{kk,\max}} \end{matrix}\left|\begin{matrix}\sum_{l'= 1}^{L_k}\end{matrix} \boldsymbol{\bf{g}}_{kkl'}^H[i]\boldsymbol{\bf{f}}_{kl'}^{\text{RZF}} \right|^2 \label{SRDAMRZF-3} \\ \notag
&=\begin{matrix}\sum_{i=\Delta_{kk,\min},i\ne 0}^{\Delta_{kk,\max}} \end{matrix}\left|\begin{matrix}\sum_{l'= 1}^{L_k}\end{matrix} \sqrt{p_{kl'}}e^{j\phi_{kl'}}\boldsymbol{\bf{g}}_{kkl'}^H[i]{\tilde{\boldsymbol{\bf{f}}}_{kl'}}/{\|\tilde{\boldsymbol{\bf{f}}}_{kl'}\|} \right|^2\\ \notag
 &=\begin{matrix}\sum_{i=\Delta_{kk,\min},i\ne 0}^{\Delta_{kk,\max}}\end{matrix} |\boldsymbol{\bf{a}}_k^H\boldsymbol{\bf{u}}_{kk}[i]|^2=\|\boldsymbol{\bf{a}}_k^H\boldsymbol{\bf{U}}_{kk}\|^2,
\end{align}
where  $\boldsymbol{\bf{u}}_{kk}[i]=\left[\boldsymbol{\bf{g}}_{kk1}^H[i]\frac{\tilde{\boldsymbol{\bf{f}}}_{k1}}{\|\tilde{\boldsymbol{\bf{f}}}_{k1}\|},..., \boldsymbol{\bf{g}}_{kkL_k}^H[i]\frac{\tilde{\boldsymbol{\bf{f}}}_{kL_k}}{\|\tilde{\boldsymbol{\bf{f}}}_{kL_k}\|}\right]^H$,  and $\boldsymbol{\bf{U}}_{kk}=\big[\boldsymbol{\bf{u}}_{kk}[\Delta_{kk,\min}] ,...,  \boldsymbol{\bf{u}}_{kk}[\Delta_{kk,\max}]\big]\in \mathbb{C}^{L_k \times (\Delta_{kk,\text{span}}-1)}$.  Following the similar definitions, the IUI power in the denominator of \eqref{RRDAM-3} is written as $\sum_{k'\ne k}^{K} \|\boldsymbol{\bf{a}}_{k'}^H{\boldsymbol{\bf{U}}}_{kk'}\|^2$. Thus, the SINR in \eqref{RRDAM-3} with the per-path-based RZF beamforming becomes
\begin{equation}
\gamma_k^{\text{RZF}}=\frac{|\boldsymbol{\bf{a}}_k^H\boldsymbol{\bf{u}}_k|^2}{\|\boldsymbol{\bf{a}}_k^H\boldsymbol{\bf{U}}_{kk}\|^2 +\sum_{k'\ne k}^{K} \|\boldsymbol{\bf{a}}_{k'}^H{\boldsymbol{\bf{U}}}_{kk'}\|^2+\sigma^2 }. \label{SRDAMRZF-4}
\end{equation}

The sum rate can be maximized by optimizing the  vectors $\{\boldsymbol{\bf{a}}_k\}_{k=1}^K$ via solving the following problem
\begin{align}
\max\limits_{\{\boldsymbol{\bf{a}}_k\}_{k=1}^K} ~&\sum_{k=1}^K \text{log}_2\left(1+ \frac{|\boldsymbol{\bf{a}}_k^H\boldsymbol{\bf{u}}_k|^2}{\|\boldsymbol{\bf{a}}_k^H\boldsymbol{\bf{U}}_{kk}\|^2 +\sum_{k'\ne k}^{K} \|\boldsymbol{\bf{a}}_{k'}^H{\boldsymbol{\bf{U}}}_{kk'}\|^2+\sigma^2 } \right)\label{SRDAMRZF-5}    \nonumber \\ 
\text{ s.t.}~ &\begin{matrix}\sum_{k=1}^K\end{matrix} \|\boldsymbol{\bf{a}}_k\|^2\leq P.
\end{align}
The above problem is non-convex, which cannot be directly solved. By introducing the slack variables $\{\bar{\gamma}_k\}$, problem \eqref{SRDAMRZF-5} can be transformed into
\begin{align}
&\max\limits_{\{\boldsymbol{\bf{a}}_k, \bar{\gamma}_k\}_{k=1}^K} \sum_{k=1}^K \text{log}_2(1+ \bar{\gamma}_k  )\label{SRDAMRZF-6}  \\ \notag
\text{ s.t.}~&\|\boldsymbol{\bf{a}}_k^H\boldsymbol{\bf{U}}_{kk}\|^2 +\sum_{k'\ne k}^{K} \|\boldsymbol{\bf{a}}_{k'}^H{\boldsymbol{\bf{U}}}_{kk'}\|^2+\sigma^2\leq \frac{|\boldsymbol{\bf{a}}_k^H\boldsymbol{\bf{u}}_k|^2}{\bar{\gamma}_k}, \forall k,\\  \notag
& \sum_{k=1}^K \|\boldsymbol{\bf{a}}_k\|^2\leq P, \\ \notag
& \bar{\gamma}_k\geq 0, \forall k.
\end{align}
Though problem \eqref{SRDAMRZF-6} is still non-convex, an efficient locally optimal solution can be obtained by using SCA technique.  Specifically, the right-hand-side of the first constraint  is quadratic-over-linear, which is convex and thus is globally lower bounded by its first-order Taylor expansion, i.e., 
\begin{equation}
\frac{|\boldsymbol{\bf{a}}_k^H\boldsymbol{\bf{u}}_k|^2}{\bar{\gamma}_k}\geq\frac{1}{\bar{\gamma}_k^{(r)}}|(\boldsymbol{\bf{a}}_k^{(r)})^H\boldsymbol{\bf{u}}_k|^2 + \Re \{\nabla f(\boldsymbol{\bf{a}}_k^{(r)},\bar{\gamma}_k^{(r)})^H\boldsymbol{\bf{e}}_k\}, \label{SRDAMRZF-7}
\end{equation}
where $\boldsymbol{\bf{a}}_k^{(r)}$ and $\bar{\gamma}_k^{(r)}$ denote the resulting solution at the $r$-th iteration, $\nabla f(\boldsymbol{\bf{a}}_k^{(r)},\bar{\gamma}_k^{(r)})=\Big[(\frac{2}{\bar{\gamma}_k^{(r)}}\boldsymbol{\bf{u}}_{k}\boldsymbol{\bf{u}}_{k}^H{\boldsymbol{\bf{a}}_k^{(r)}})^H, ~\frac{-1}{(\bar{\gamma}_k^{(r)})^2}|(\boldsymbol{\bf{a}}_k^{(r)})^H\boldsymbol{\bf{u}}_{k}|^2\Big]^H$ is the gradient, and $\boldsymbol{\bf{e}}_k=[(\boldsymbol{\bf{a}}_k-\boldsymbol{\bf{a}}_k^{(r)})^H, ~\bar{\gamma}_k-\bar{\gamma}_k^{(r)}]^H$.
Therefore, for given $\boldsymbol{\bf{a}}_k^{(r)}$ and $\bar{\gamma}_k^{(r)}$ at the $r$-th iteration, the optimal value of problem \eqref{SRDAMRZF-6} is lower-bounded by that of the following problem
\begin{align}
\max\limits_{\{\boldsymbol{\bf{a}}_k, \bar{\gamma}_k\}_{k=1}^K}~& \sum_{k=1}^K \text{log}_2(1+ \bar{\gamma}_k  )\label{SRDAMRZF-8}  \\  \notag
\text{ s.t.}~& \|\boldsymbol{\bf{a}}_k^H\boldsymbol{\bf{U}}_{kk}\|^2 +\sum_{k'\ne k}^{K} \|\boldsymbol{\bf{a}}_{k'}^H{\boldsymbol{\bf{U}}}_{kk'}\|^2+\sigma^2\\ \notag
& \leq \frac{1}{\bar{\gamma}_k^{(r)}}|(\boldsymbol{\bf{a}}_k^{(r)})^H\boldsymbol{\bf{u}}_k|^2 +  \Re\{\nabla f(\boldsymbol{\bf{a}}_k^{(r)},\bar{\gamma}_k^{(r)})^H\boldsymbol{\bf{e}}_k\},\\ \notag
& \sum_{k=1}^K \|\boldsymbol{\bf{a}}_k\|^2\leq P, \\  \notag
&\bar{\gamma}_k\geq 0, \forall k.
\end{align}
Problem \eqref{SRDAMRZF-8} is convex, which can be efficiently solved by the standard convex optimization toolbox, such as CVX. By successively updating the local point $\{\boldsymbol{\bf{a}}_k^{(r)}, \bar{\gamma}_k^{(r)}\}_{k=1}^K$, an efficient solution to  problem \eqref{SRDAMRZF-6} can be obtained. The details for solving problem \eqref{SRDAMRZF-6} are  summarized in Algorithm \ref{sca}. 
Note that since the resulting objective value of problem \eqref{SRDAMRZF-6} is non-decreasing over each iteration, Algorithm \ref{sca} is guaranteed to converge.

 \begin{algorithm}[t]
  \caption{SCA-based  Optimization for Per-Path-Based RZF Beamforming}
  \label{sca}
  \begin{algorithmic}[1]
   \State Initialize a feasible solution $\{\boldsymbol{\bf{a}}_k^{(0)}, \bar{\gamma}_k^{(0)}\}_{k=1}^K$ to problem \eqref{SRDAMRZF-6}. Let $r=0$.
   \Repeat
   \State {Solve the convex optimization  problem \eqref{SRDAMRZF-8} for given }
   \Statex\hspace{5.5mm}{$\{\boldsymbol{\bf{a}}_k^{(r)}, \bar{\gamma}_k^{(r)}\}$, and denote the optimal solution as
    \Statex\hspace{5.5mm}$\{\boldsymbol{\bf{a}}_k^{(r+1)}, \bar{\gamma}_k^{(r+1)}\}$. }
    \State Update $r=r+1$.
    \Until{ The  fractional increase of objective value of problem \eqref{SRDAMRZF-6}  is below a certain threshold}.
  \end{algorithmic}
\end{algorithm}

In the following, we consider the achievable sum rate of the benchmarking schemes.

\subsection{Achievable Sum Rate by the Strongest-Path-Based Beamforming} 

For the transmitted signal in \eqref{RRSP-1}, we analyze the strongest-path-based MRT, ZF, and RZF beamforming, respectively.

\subsubsection{MRT Beamforming}

The strongest-path-based MRT beamforming scheme is given by
\begin{equation}
\boldsymbol{\bf{f}}_{k}^{\text{SP-MRT}}=\sqrt{P}{\boldsymbol{\bf{h}}_{k1}}/{\|\bar{\boldsymbol{\bf{H}}}\|_{F}}\label{SRSPMRT-1},
 \end{equation}
with $\bar{\boldsymbol{\bf{H}}}=[\boldsymbol{\bf{h}}_{11},..,\boldsymbol{\bf{h}}_{K1}]$. The resulting SINR of UE $k$ is expressed as 
\begin{equation}
\begin{split}
&\gamma_{k}^{\text{SP-MRT}}=\label{SRSPMRT-2} \\
&\frac{\|\boldsymbol{\bf{h}}_{k1} \|^4}{ \sum_{l\ne 1}^{L_k} |\boldsymbol{\bf{h}}_{kl}^H\boldsymbol{\bf{h}}_{k1} |^2 +   \sum_{k'\ne k}^K\sum_{l=1}^{L_{k}}|\boldsymbol{\bf{h}}_{kl}^H \boldsymbol{\bf{h}}_{k'1}|^2+ \|\bar{\boldsymbol{\bf{H}}}\|_{F}^2\sigma^2/P },
\end{split}
\end{equation}
and  the sum rate of the strongest-path-based MRT beamforming is  $\sum_{k=1}^K\text{log}_2(1+\gamma_{k}^{\text{SP-MRT}})$.

\subsubsection{ZF Beamforming} \label{SRSPZF}

For the strongest-path-based beamforming schemes, the ZF beamforming vectors  are designed so that the ISI and IUI in \eqref{RRSP-2} are all eliminated, i.e., 
\begin{align}
\boldsymbol{\bf{h}}_{kl}^H \boldsymbol{\bf{f}}_{k}^{\text{SP-ZF}}&=0, ~\forall l\neq 1, \forall k, \label{SRSPZF-1} \\ 
\boldsymbol{\bf{h}}_{kl}^H \boldsymbol{\bf{f}}_{k'}^{\text{SP-ZF}}&=0, ~\forall k'\neq k, \text{and} ~\forall l. \label{SRSPZF-2}
\end{align}
Denote by $\boldsymbol{\bf{H}}_{k1}\in \mathbb{C}^{M_t \times (L_{\mathrm{tot}}-1)}$  the submatrix of $\boldsymbol{\bf{H}}$   excluding the column $\boldsymbol{\bf{h}}_{k1}$. Thus, the ZF constraints in \eqref{SRSPZF-1} and \eqref{SRSPZF-2} is compactly written as
\begin{equation}
\boldsymbol{\bf{H}}_{k1}^H \boldsymbol{\bf{f}}_{k}^{\text{SP-ZF}}=\boldsymbol{\bf{0}}_{(L_{\mathrm{tot}}-1)\times 1}, ~\forall k. \\ \label{SRSPZF-3}
\end{equation}
The above ZF constraint is feasible when $M_t\geq L_{\mathrm{tot}}$.

Similar to Section \ref{ZF}, let $\boldsymbol{\bf{F}}^{\text{SP-ZF}}=[\boldsymbol{\bf{f}}_{1}^{\text{SP-ZF}},...,\boldsymbol{\bf{f}}_{K}^{\text{SP-ZF}}]\in \mathbb{C}^{M_t \times K}$, where $\boldsymbol{\bf{F}}^{\text{SP-ZF}}=\boldsymbol{\bf{W}}^{\text{SP-ZF}}\boldsymbol{\bf{\hat{V}}}^{\frac{1}{2}}$, with $\boldsymbol{\bf{W}}^{\text{SP-ZF}}=[\boldsymbol{\bf{w}}_{11}^{\text{SP-ZF}},...,\boldsymbol{\bf{w}}_{K1}^{\text{SP-ZF}}]\in \mathbb{C}^{M_t\times K}$ and power allocation matrix $\boldsymbol{\bf{\hat{V}}}^{\frac{1}{2}}=\text{diag}\{\hat{v}_{1},...,\hat{v}_{K}\}\in \mathbb{C}^{K \times K}$. 
As a result, the ZF beamforming $\boldsymbol{\bf{f}}_{k}^{\text{SP-ZF}}$ is expressed as $\boldsymbol{\bf{f}}_{k}^{\text{SP-ZF}}=\sqrt{\hat{v}_{k}}{\boldsymbol{\bf{w}}_{k1}}$, where  ${\boldsymbol{\bf{w}}}_{k1}$ is obtained by taking the $\big(\sum_{j=1}^{k-1}L_j+1\big)$th column of $(\boldsymbol{\bf{H}}^H)^{\dagger}$. By substituting $\boldsymbol{\bf{f}}_{k}^{\text{SP-ZF}}$ into \eqref{RRSP-2}, the received signal reduces to
\begin{align}
y_{k, \text{SP}}[n] &=\sqrt{\hat{v}_{k}} \boldsymbol{\bf{h}}_{k1}^H {\boldsymbol{\bf{w}}_{k1}} s_k[n-n_{k1}]+z_k[n] \label{SRSPZF-4} \\ \notag
&=\sqrt{\hat{v}_{k}}  s_k[n-n_{k1}]+z_k[n].
\end{align}
The sum rate maximization problem is  formulated as              
\begin{equation}
\begin{split}
\max\limits_{\{\hat{v}_k\}_{k=1}^K} &\sum_{k=1}^K \text{log}_2\left(1+\frac{\hat{v}_{k}}{ \sigma^2}\right)\label{SRSPZF-4} \\ 
\text{s.t.}~&\sum_{k=1}^K \hat{v}_{k}\|\boldsymbol{\bf{w}}_{k1}\|^2 \leq P.
\end{split}
\end{equation}
By defining $\tilde{v}_k=\hat{v}_{k}\|\boldsymbol{\bf{w}}_{k1}\|^2$, problem \eqref{SRSPZF-4} is equivalently re-expressed as
\begin{equation}
\begin{split}
\max\limits_{\{\tilde{v}_k\}_{k=1}^K} &\sum_{k=1}^K \text{log}_2\left(1+\frac{\tilde{v}_{k}}{ \|\boldsymbol{\bf{w}}_{k1}\|^2\sigma^2}\right)\label{SRSPZF-5} \\ 
\text{s.t.}~&\sum_{k=1}^K  \tilde{v}_k \leq P, \\
&\tilde{v}_k \geq 0, \forall k.
\end{split}
\end{equation}
Thus,  the optimal power allocation coefficient can be obtained  by WF strategy, and the accordingly achieved sum rate is also obtained.

\subsubsection{RZF Beamforming} \label{SP-RZF}

The  strongest-path-based RZF beamforming  is given by
\begin{equation}
\boldsymbol{\bf{f}}_{k}^{\text{SP-RZF}}=\sqrt{\hat{p}_{k}}e^{j\hat{\phi}_k}{\tilde{\boldsymbol{\bf{f}}}_{k1}}/{\|\tilde{\boldsymbol{\bf{f}}}_{k1}\|} \label{SRSPRZF-1},
\end{equation}
 where $\hat{p}_k$ and $e^{j\hat{\phi}_k}$ denote the power allocation and the phase rotation of  UE $k$, respectively, and $\tilde{\boldsymbol{\bf{f}}}_{k1}$ is obtained by taking the $\big(\sum_{j=1}^{k-1}L_j+1\big)$th column of the matrix $\tilde{\boldsymbol{\bf{F}}}$. By substituting  \eqref{SRSPRZF-1} into \eqref{RRSP-3} and after some manipilations, the sum rate for the strongest-path-based RZF beamforming  is given by
\begin{align}
&R_{\text{SP-RZF}}=\label{SRSPRZF-2} \\ \notag
&\sum_{k=1}^K\text{log}_2\left(1+\frac{\hat{p}_k b_{kk,1}}{\hat{p}_k \sum_{l\ne 1}^{L_k} b_{kk,l}+  \sum_{k'\ne k}^K \hat{p}_{k'}\sum_{l=1}^{L_{k}} b_{kk',l}+\sigma^2}\right)\\ \notag
&=\sum_{k=1}^K\text{log}_2\left(\frac{\sum_{k'=1}^K \hat{p}_{k'}\sum_{l=1}^{L_{k}} b_{kk',l}+\sigma^2}{\hat{p}_k \sum_{l\ne 1}^{L_k} b_{kk,l}+  \sum_{k'\ne k}^K \hat{p}_{k'}\sum_{l=1}^{L_{k}} b_{kk',l}+\sigma^2}\right)\\ \notag
&=\sum_{k=1}^K\text{log}_2\left(\frac{\boldsymbol{\bf{p}}^T\boldsymbol{\bf{b}}_{1,k}+\sigma^2}{\boldsymbol{\bf{p}}^T\boldsymbol{\bf{b}}_{2,k}+\sigma^2}\right), 
\end{align}
where  $b_{kk',l}=\left|\boldsymbol{\bf{h}}_{kl}^H \frac{\tilde{\boldsymbol{\bf{f}}}_{k'1}}{\|\tilde{\boldsymbol{\bf{f}}}_{k'1}\|}\right|^2$,  $\boldsymbol{\bf{p}}=[\hat{p}_1,...,\hat{p}_K]^T\in \mathbb{C}^{K \times 1}$, $\boldsymbol{\bf{b}}_{1,k}=\left[\sum_{l=1}^{L_{k}} b_{k1,l},...,\sum_{l=1}^{L_{k}} b_{kK,l}\right]^T\in \mathbb{R}^{K \times 1}$, and $\boldsymbol{\bf{b}}_{2,k} =\Big[\sum_{l=1}^{L_{k}} b_{k1,l},...,\sum_{l=1}^{L_{k}} b_{k(k-1),l},\sum_{l\ne 1}^{L_k} b_{kk,l}, \sum_{l=1}^{L_{k}} b_{k(k+1),l},..,\\
\sum_{l=1}^{L_{k}} b_{kK,l}\Big]^T\in \mathbb{C}^{K \times 1}$. It is observed that different from DAM transmission, the phase rotation  has no impact on the resulting sum rate. Thus, the sum rate maximization of the strongest-path-based RZF beamforming can be formulated as 
 \begin{align}
\max\limits_{\boldsymbol{\bf{p}}} &\sum_{k=1}^K \text{log}_2\left(\boldsymbol{\bf{p}}^T\boldsymbol{\bf{b}}_{1,k}+\sigma^2\right) -\sum_{k=1}^K\text{log}_2\left(\boldsymbol{\bf{p}}^T\boldsymbol{\bf{b}}_{2,k}+\sigma^2\right) 
 \label{SRSPRZF-3} \nonumber \\
\text{s.t.}~& \boldsymbol{\bf{1}}_K^T \boldsymbol{\bf{p}} \leq P,\\ 
&\boldsymbol{\bf{p}} \geq  \boldsymbol{0}_{K}.\nonumber
\end{align}

Note that problem \eqref{SRSPRZF-3} is non-convex since  $\text{log}_2(\boldsymbol{\bf{p}}^T\boldsymbol{\bf{b}}_{2,k}+\sigma^2)$  is a concave  function with $\boldsymbol{\bf{p}}$. By applying the first-order Taylor expansion at any given point $\boldsymbol{\bf{p}}^{(r)}$, $\text{log}_2(\boldsymbol{\bf{p}}^T\boldsymbol{\bf{b}}_{2,k}+\sigma^2) $ is upper bounded by
 \begin{align}
&\text{log}_2(\boldsymbol{\bf{p}}^T\boldsymbol{\bf{b}}_{2,k}+\sigma^2) \leq \text{log}_2\left((\boldsymbol{\bf{p}}^{(r)})^T\boldsymbol{\bf{b}}_{2,k}+\sigma^2\right) \label{SRSPRZF-4} \notag \\ 
&+ \frac{\boldsymbol{\bf{b}}_{2,k}^T(\boldsymbol{\bf{p}}-\boldsymbol{\bf{p}}^{(r)})}{(\boldsymbol{\bf{p}}^{(r)})^T\boldsymbol{\bf{b}}_{2,k}+\sigma^2} \text{log}_2e.
\end{align}

With the upper bound  \eqref{SRSPRZF-4}, problem \eqref{SRSPRZF-3} can be transformed into
\begin{align}
\max\limits_{\boldsymbol{\bf{p}}} ~&\sum_{k=1}^K \text{log}_2(\boldsymbol{\bf{p}}^T\boldsymbol{\bf{b}}_{1,k}+\sigma^2) -\sum_{k=1}^K\text{log}_2\left((\boldsymbol{\bf{p}}^{(r)})^T\boldsymbol{\bf{b}}_{2,k}+\sigma^2\right)
\label{SRSPRZF-5} \notag \\
&- \sum_{k=1}^K\frac{\boldsymbol{\bf{b}}_{2,k}^T(\boldsymbol{\bf{p}}-\boldsymbol{\bf{p}}^{(r)})}{(\boldsymbol{\bf{p}}^{(r)})^T\boldsymbol{\bf{b}}_{2,k}+\sigma^2} \text{log}_2e\\ \notag
\text{s.t.}~&\boldsymbol{\bf{1}}_K^T \boldsymbol{\bf{p}}  \leq P,\\  \notag
&\boldsymbol{\bf{p}} \geq  \boldsymbol{0}_{K}.
\end{align}
The details for solving problem \eqref{SRSPRZF-5} are similar to Algorithm \ref{sca}, which is omitted for brevity.

\subsection{Achievable Sum Rate by OFDM} 

For the transmitted signal in \eqref{RROFDM-2}, we analyze the OFDM-based MRT, ZF, and RZF beamforming, respectively.

\subsubsection{MRT Beamforming}

Denote by $\boldsymbol{\bf{\hat{H}}}=[\boldsymbol{\bf{{H}}}_1,...,\boldsymbol{\bf{{H}}}_M]\in \mathbb{C}^{M_t\times KM}$ the frequency-domain channel matrix, where  $\boldsymbol{\bf{{H}}}_m=[\boldsymbol{\bf{h}}_{1,m},...,\boldsymbol{\bf{h}}_{K,m}] \in \mathbb{C}^{M_t\times K}$. The MRT beamforming for OFDM transmission is
\begin{equation}
 \boldsymbol{\bf{d}}^{\text{MRT}}_{k,m}=\sqrt{P_s}{\boldsymbol{\bf{h}}_{k,m}}/{\|\boldsymbol{\bf{\hat{H}}}\|_F},\label{SROFDMMRT-1} 
 \end{equation}
where $P_s=MP$. By substituting \eqref{SROFDMMRT-1} into \eqref{RROFDM-5}, the resulting SINR is given by
\begin{equation}
\gamma_{k,m}^{\text{OFDM-MRT}}=\frac{\|\boldsymbol{\bf{h}}_{k,m}\|^4}{\sum_{k'\ne k}^K |\boldsymbol{\bf{h}}_{k,m}^H\boldsymbol{\bf{h}}_{k',m}|^2+\|\boldsymbol{\bf{\hat{H}}}\|_F^2\hat{\sigma}^2/P_s}. \label{SROFDMMRT-2}     %~\eqref{111}
\end{equation}
%\vspace{-0.3cm}
The sum rate of MRT Beamforming for OFDM without considering CP overhead is $ \frac{1}{M}\sum_{m=1}^M\sum_{k=1}^K\text{log}_2(1+\gamma_{k,m}^{\text{OFDM-MRT}})$.

\subsubsection{ZF Beamforming} 
For ZF beamforming, $\boldsymbol{\bf{d}}_{k,m}, k = 1,...,K, m=1,...,M,  $ are designed so that the IUI of each sub-carrier $m$ in \eqref{RROFDM-4} is eliminated, i.e., 
\begin{equation}
\boldsymbol{\bf{h}}_{k,m}^H \boldsymbol{\bf{d}}_{k',m}^{\text{ZF}}=0, ~\forall k'\neq k, m. \\ \label{SROFDMZF-1}
\end{equation}
Let $\boldsymbol{\bf{H}}_{k,m}=[\boldsymbol{\bf{h}}_{1,m},...,\boldsymbol{\bf{h}}_{k-1,m},\boldsymbol{\bf{h}}_{k+1,m},\boldsymbol{\bf{h}}_{K,m}]\in \mathbb{C}^{M_t \times (K-1)}$. Thus, the sub-carrier based ZF constraints in \eqref{SROFDMZF-1} can be compactly written as
\begin{equation}
\boldsymbol{\bf{H}}_{k,m}^H \boldsymbol{\bf{d}}_{k,m}^{\text{ZF}}=\boldsymbol{\bf{0}}_{(K-1)\times 1}, ~\forall k, m. \\ \label{SROFDMZF-2}
\end{equation}
The above ZF constraint is feasible when $M_t\geq K$. The sub-carrier based ZF beamforming matrix is obtained similarly as the per-path-based ZF beamforming in Section \ref{ZF}, given by $\boldsymbol{\bf{d}}_{k,m}^{\text{ZF}} =\sqrt{\lambda_{k,m}}\boldsymbol{\bf{b}}_{k,m}$, where $\lambda_{k,m}$ denotes the power allocation coefficient, $\boldsymbol{\bf{b}}_{k,m}$ is the $k$th column of the matrix $ {\boldsymbol{\bf{H}}}_m({\boldsymbol{\bf{H}}}_m^H{\boldsymbol{\bf{H}}}_m)^{-1}$. By substituting $\boldsymbol{\bf{d}}_{k,m}^{\text{ZF}}$ into \eqref{RROFDM-4}, the received signal reduces to 
\begin{align} %\begin{matrix}
y_{k,m}&=\sqrt{\lambda_{k,m}}\boldsymbol{\bf{h}}_{k,m}^H\boldsymbol{\bf{b}}_{k,m} s_{k,m} + {z}_{k,m}\\ \notag
&=\sqrt{\lambda_{k,m}}s_{k,m} + {z}_{k,m}. \label{SROFDMZF-3}
\end{align}
The SNR of the $m$th sub-carrier for UE $k$  is $\gamma_{k,m}^{\text{ZF}}={\lambda_{k,m} }/{ \hat{\sigma}^2}$. Thus, the optimal power allocation coefficients $\lambda_{k,m}, \forall k,m$, to  maximize the sum rate can be found by solving the following problem
\begin{equation}
\begin{split}
\max\limits_{{\lambda}_{k,m}, \forall k,m}~ &\frac{1}{M}\sum_{m=1}^M\sum_{k=1}^K  \text{log}_2\left(1+\frac{\lambda_{k,m}}{\hat{\sigma}^2}\right)\label{SROFDMZF-4}\\
\text{s.t.}~&\begin{matrix}\sum_{m=1}^M\end{matrix}\begin{matrix}\sum_{k=1}^K\end{matrix} \lambda_{k,m}\|\boldsymbol{\bf{b}}_{k,m}\|^2 \leq MP,\\
& {\lambda}_{k,m}\geq 0, \forall k,m,
\end{split}
\end{equation}
By defining $c_{k,m}=\lambda_{k,m}\|\boldsymbol{\bf{b}}_{k,m}\|^2$, problem \eqref{SROFDMZF-4} can be equivalently written as
\begin{equation}
\begin{split}
\max\limits_{c_{k,m}, \forall k,m}~ &\frac{1}{M} \sum_{m=1}^M \sum_{k=1}^K\text{log}_2\left(1+\frac{c_{k,m}}{\hat{\sigma}^2\|\boldsymbol{\bf{b}}_{k,m}\|^2}\right)\label{SROFDMZF-5}\\
\text{s.t.}~&\begin{matrix}\sum_{m=1}^M\end{matrix}\begin{matrix}\sum_{k=1}^K\end{matrix}  c_{k,m} \leq MP, \\
& c_{k,m} \geq 0, \forall k,m.
\end{split}
\end{equation}
The optimal solution to problem \eqref{SROFDMZF-5} can be obtained  by the classical WF solution and the sum rate of ZF beamforming for OFDM is obtained accordingly.

\subsubsection{RZF Beamforming} 

Let ${\boldsymbol{\bf{D}}}_m=\boldsymbol{\bf{{H}}}_m(\boldsymbol{\bf{{H}}}_m^H\boldsymbol{\bf{{H}}}_m+\epsilon_{\text{OFDM}}\boldsymbol{\bf{{I}}}_{K} )^{-1}=[\bar{\boldsymbol{\bf{d}}}_{1m},...,\bar{\boldsymbol{\bf{d}}}_{Km}]$ with $\epsilon_{\text{OFDM}}=K\hat{\sigma}^2/P$ \cite{RZF}. Thus, the sub-carrier based RZF beamforming can be given by
\begin{equation}
{\boldsymbol{\bf{d}}}_{k,m}^{\text{RZF}} =  \sqrt{p_{k,m}}e^{j{\phi}_{k,m}}{\bar{\boldsymbol{\bf{d}}}_{k,m}}/{\|\bar{\boldsymbol{\bf{d}}}_{k,m}\|},\label{SROFDMRZF-1}
\end{equation}
where $p_{k,m}$ and $e^{j{\phi}_{k,m}}$ denote the power allocation and the phase rotation of UE $k$ at sub-carrier $m$,  respectively.
By substituting \eqref{SROFDMRZF-1} into \eqref{RROFDM-5}, the sum rate of RZF beamforming for OFDM can be written as
\begin{align}
R_{\text{OFDM-RZF}}&=  \frac{1}{M}\sum_{m=1}^M\sum_{k=1}^K \text{log}_2\left(1+\frac{p_{k,m}a_{kk, m}}{\sum_{k'\ne k}^K p_{k',m}a_{kk', m}+\hat{\sigma}^2}\right) \label{SROFDMRZF-2}  \nonumber  \\ 
&=\frac{1}{M}\sum_{m=1}^M\sum_{k=1}^K \text{log}_2\left(\frac{\sum_{k'=1}^K p_{k',m}a_{kk', m}+\hat{\sigma}^2}{\sum_{k'\ne k}^K p_{k',m}a_{kk', m}+\hat{\sigma}^2}\right)\nonumber \\  
&=\frac{1}{M}\sum_{m=1}^M\sum_{k=1}^K \text{log}_2\left(\frac{\boldsymbol{\bf{p}}_{m}^T{\boldsymbol{\bf{a}}}_{k,m,1}+\hat{\sigma}^2}{\boldsymbol{\bf{p}}_{m}^T{\boldsymbol{\bf{a}}}_{k,m,2}+\hat{\sigma}^2}\right),
\end{align}
where $a_{kk',m}=\left|\boldsymbol{\bf{h}}_{k,m}^H\frac{{\boldsymbol{\bf{d}}}_{k',m}}{\|{\boldsymbol{\bf{d}}}_{k',m}\|}\right|^2$, ${\boldsymbol{\bf{p}}}_{m}=[{p}_{1,m},...,p_{K,m}]^T$, $\boldsymbol{\bf{a}}_{k,m,1}=[a_{k1,m},...,a_{kK,m}]^T$, and $\boldsymbol{\bf{a}}_{k,m,2}=[a_{k1,m},...,a_{k(k-1),m},0,a_{k(k+1),m},...,a_{kK,m}]^T$.
 
 As a result, the sum rate maximization for OFDM is equivalently formulated as (by discarding the constant term $\frac{1}{M}$)
\begin{align}
\max\limits_{\boldsymbol{\bf{p}}_m, \forall m} ~&\sum_{m=1}^M\sum_{k=1}^K \text{log}_2(\boldsymbol{\bf{p}}_{m}^T{\boldsymbol{\bf{a}}}_{k,m,1}+\hat{\sigma}^2)-  \nonumber \\ 
&\sum_{m=1}^M\sum_{k=1}^K\text{log}_2(\boldsymbol{\bf{p}}_{m}^T{\boldsymbol{\bf{a}}}_{k,m,2}+\hat{\sigma}^2)\label{SROFDMRZF-3}    \nonumber \\ 
\text{ s.t.}~ &\begin{matrix}\sum_{m=1}^M\end{matrix}\begin{matrix}\sum_{k=1}^K\end{matrix} p_{k,m}\leq MP, \nonumber \\ 
 &p_{k,m} \geq 0, \forall k,m.
\end{align}

Similar to \ref{SP-RZF}, problem \eqref{SROFDMRZF-3} is solved via the SCA technique.

\section{Guard Interval and PAPR Analysis}

In this section, we discuss the guard interval overhead and PAPR for single-carrier DAM, the strongest-path-based beamforming and OFDM.

\subsection{Guard Interval Overhead}
Denote by $G_c= T_c/T_s$ the number of signal samples  within each channel coherence time,  where $T_s$ denotes the single-carrier symbol interval that is inversely proportional to bandwidth,  and $T_c$ represents the channel coherence time.  To avoid the inter-block interference, the single-carrier  schemes including  multi-user DAM and strongest-path-based beamforming only need a guard interval $G_{sc}$ at the beginning  of each channel coherence block \cite{DAM-OFDM}. However, a CP of length $G_{\text{CP}}$ is required for each OFDM symbol. Thus, each OFDM symbol duration is $(M+G_{\text{CP}})T_s$, and the number of OFDM symbols in each channel coherence time is $n_{\text{OFDM}}=G_c/(M+G_{\text{CP}})$.
More specifically,  the guard interval of DAM is $G_{\text{DAM}}\approx 2n_{\text{max}}$ with $n_{\text{max}}=\text{max}\{n_{\text{1,max}},...,n_{K,\text{max}}\}$ \cite{DAM}, \cite{DAM-OFDM}, while the CP length for OFDM  and guard interval of  SP scheme are $G_{\text{CP}}=n_{\text{max}}$ and $G_{\text{SP}}=n_{\text{max}}$, respectively. Then  the guard interval overhead of DAM, OFDM, and SP scheme are $2n_{\text{max}}/G_c$, $n_{\text{OFDM}}G_{\text{CP}}/G_c=n_{\text{max}}/(n_{\text{max}}+M)$, and $n_{\text{max}}/G_c$, respectively, where DAM has a significantly reduction compared with OFDM when $n_{\text{OFDM}}\gg2$. Therefore, after considering the guard interval overhead, the effective spectral efficiency of OFDM and   single-carrier transmission in bps/Hz are
\begin{align}
R_{\text{OFDM}} &=\frac{1}{M+G_{\text{CP}}}\sum_{k=1}^K\sum_{m=1}^M \text{log}_2(1+\gamma_{k,m}) \label{GI-1},
\end{align} 
and
\begin{equation}
R_{\text{SC}} = \frac{G_c- G_{\text{SC}}}{G_c}\sum_{k=1}^K\text{log}_2(1+\gamma_k), \label{GI-2}
\end{equation} 
respectively, where  $\text{SC}\in \{\text{DAM}, \text{SP}\}$, $\gamma_{k,m}$ denotes the SINR/SNR of UE $k$ at $m$th sub-carrier for OFDM transmission, and $\gamma_k$ is the SINR/SNR of UE $k$ for single-carrier transmission.

%\begin{figure}[t]
%\centering
	%\includegraphics[scale=0.6]{block.jpg}
%\caption{An illustration of multi-user  DAM or strongest path and OFDM block structures.}\label{fig66}
 %\vspace{-3ex}
%\end{figure} 

\subsection{PAPR}

In this subsection, we analyze the PAPR for multi-user DAM, the strongest-path-based beamforming, and OFDM.

\subsubsection{Multi-User DAM}
For  multi-user DAM, it is observed from \eqref{2} that the  transmit signal at the $m_t$th  antenna is 
\begin{equation}
x^{m_t}[n]=\begin{matrix}\sum_{k=1}^{K}\end{matrix}\begin{matrix}\sum_{l'=1}^{L_k}\end{matrix}f_{kl'}^{m_t} s_{k}[n-\kappa_{kl'}], 1\leq m_t\leq M_t, \label{PAPR-1}
\end{equation}
where $f_{kl'}^{m_t}$ denote the $m_t$th element of $\boldsymbol{\bf{f}}_{kl'}$, and the symbol constellation $s_{k}[n]\in \mathcal{A}$ are obtained from the quadrature amplitude modulation (QAM) alphabet $\mathcal{A}  = \{a_1 ,..., a_Q\}$ of size $Q$. Similar to \cite{DAM-ISAC}, \cite{PAPR}, the PAPR  of multi-user DAM  is defined as   
\begin{equation}
\text{PAPR}_{\text{MU-DAM}} = \max\limits_{1\leq m_t\leq M_t}\text{PAPR}_{m_t}, \label{PAPR-2}
\end{equation}
where the $\text{PAPR}_{m_t}$ is the PAPR  of the $m_t$th transmit antenna, given by 
\begin{align} %\max\limits_{1\leq n\leq N_c, \atop 1\leq m_t\leq M_t}
&\text{PAPR}_{m_t}=\max\limits_{ 1\leq n\leq G_c}\frac{ |x^{m_t}[n]|^2}{\mathbb{E}[|x^{m_t}[n]|^2]} \label{PAPR-3}\\ \notag
&=\max\limits_{ 1\leq n\leq G_c}\frac{\left|\sum_{k=1}^{K}\sum_{l'=1}^{L_k}f_{kl'}^{m_t} s_{k}[n-\kappa_{kl'}]\right|^2}{\mathbb{E}\left[\left|\sum_{k=1}^{K}\sum_{l'=1}^{L_k}f_{kl'}^{m_t} s_{k}[n-\kappa_{kl'}]\right|^2\right]}. 
\end{align}

\subsubsection{Strongest-Path-Based Beamforming}
The transmit signal of the $m_t$th transmit antenna for the strongest-path-based beamforming is
\begin{equation}
x_{\text{SP}}^{m_t}[n]=\sum_{k=1}^{K}f_k^{m_t} s_k[n]. \label{PAPR-4}
\end{equation}
 Then the PAPR of multi-user strongest-path-based beamforming is similarly defined as \eqref{PAPR-2}, with PAPR for the $m_t$th antenna given by
\begin{align} %\max\limits_{1\leq n\leq N_c, \atop 1\leq m_t\leq M_t}
\text{PAPR}_{m_t}&= \max\limits_{ 1\leq n\leq G_c}\frac{ |x_{\text{SP}}^{m_t}[n]|^2}{\mathbb{E}[|x_{\text{SP}}^{m_t}[n]|^2]} \label{PAPR-5}\\ \notag
&=\max\limits_{ 1\leq n\leq G_c} \frac{\left|\sum_{k=1}^{K}f_k^{m_t} s_k[n]\right|^2}{\mathbb{E}\left[\left|\sum_{k=1}^{K}f_k^{m_t} s_k[n]\right|^2\right]}.
\end{align}

\subsubsection{OFDM}

The time-domain transmitted signal of the $m_t$th antenna at the $n_s$th  OFDM symbol  is
\begin{align}
&X^{n_s,m_t}[n]=\frac{1}{\sqrt{M}}\sum_{m=0}^{M-1}x_m^{m_t} e^{j2\pi mn/M}\label{PAPR-51} \\ \notag 
&=\frac{1}{\sqrt{M}}\sum_{m=0}^{M-1}\sum_{k=1}^{K}d_{k,m}^{m_t} s_{k,m}^{n_s} e^{j2\pi mn/M}, 1\leq n_s \leq n_{\text{OFDM}},
\end{align}
where the symbol constellation  $s_{k,m}^{n_s}$ are same to that of multi-user DAM and strongest-path-based beamforming. Then the PAPR of the $m_t$th antenna by OFDM  is given by
\begin{align} %\max\limits_{1\leq n\leq N_c, \atop 1\leq m_t\leq M_t}
&\text{PAPR}_{m_t}= \max\limits_{\substack{ 0\leq n\leq M-1,  \\ 1 \leq n_s\leq n_{\text{OFDM}}}}\frac{ |X^{n_s,m_t}[n]|^2}{\mathbb{E}[|X^{n_s,m_t}[n]|^2]} \label{PAPR-6}\\ \notag
&=\max\limits_{\substack{ 0\leq n\leq M-1,  \\ 1 \leq n_s\leq n_{\text{OFDM}}}} \frac{\left|\sum_{m=1}^{M}\sum_{k=1}^{K}d_{k,m}^{m_t} s_{k,m}^{n_s} e^{j2\pi mn/M}\right|^2}{\mathbb{E}\left[\left|\sum_{m=1}^{M}\sum_{k=1}^{K}d_{k,m}^{m_t} s_{k,m}^{n_s} e^{j2\pi mn/M}\right|^2\right]}.
\end{align}

\section{Simulation Results} \label{Sim}

In this section, we provide simulation results to demonstrate the effectiveness of the proposed multi-user DAM scheme. We consider a mmWave system operating at $f_c=$ 28 GHz, with a total bandwidth $B=$ 128 MHz. The  noise power spectrum density is $N_0=-174$ dBm/Hz, and the  total noise power is $\sigma^2=-93$ dBm. The number of BS antennas is $M_t=128$. The channel coherence time is $T_c = 1$ ms, within which  the total number of signal samples is  $G_c = 1.28 \times 10^5$. Unless otherwise stated, the BS transmit power is $P=30$ dBm,  the number of UEs is $K =3$, and  the number of temporal-resovlable multi-paths for each UE is $L_k=L=5, \forall k$, with their discretized multi-path delays being randomly generated in $[0, 80]$. Besides, the AoDs of all the multi-paths are randomly generated in the interval $[-90^\circ,90^\circ]$, and the complex-valued gains of each path are generated based on the model developed in \cite{mmWave}. For the benchmarking scheme of OFDM, the number of sub-carriers is $M = 512$.

\begin{figure}[t]
\centering
	\includegraphics[scale=0.06]{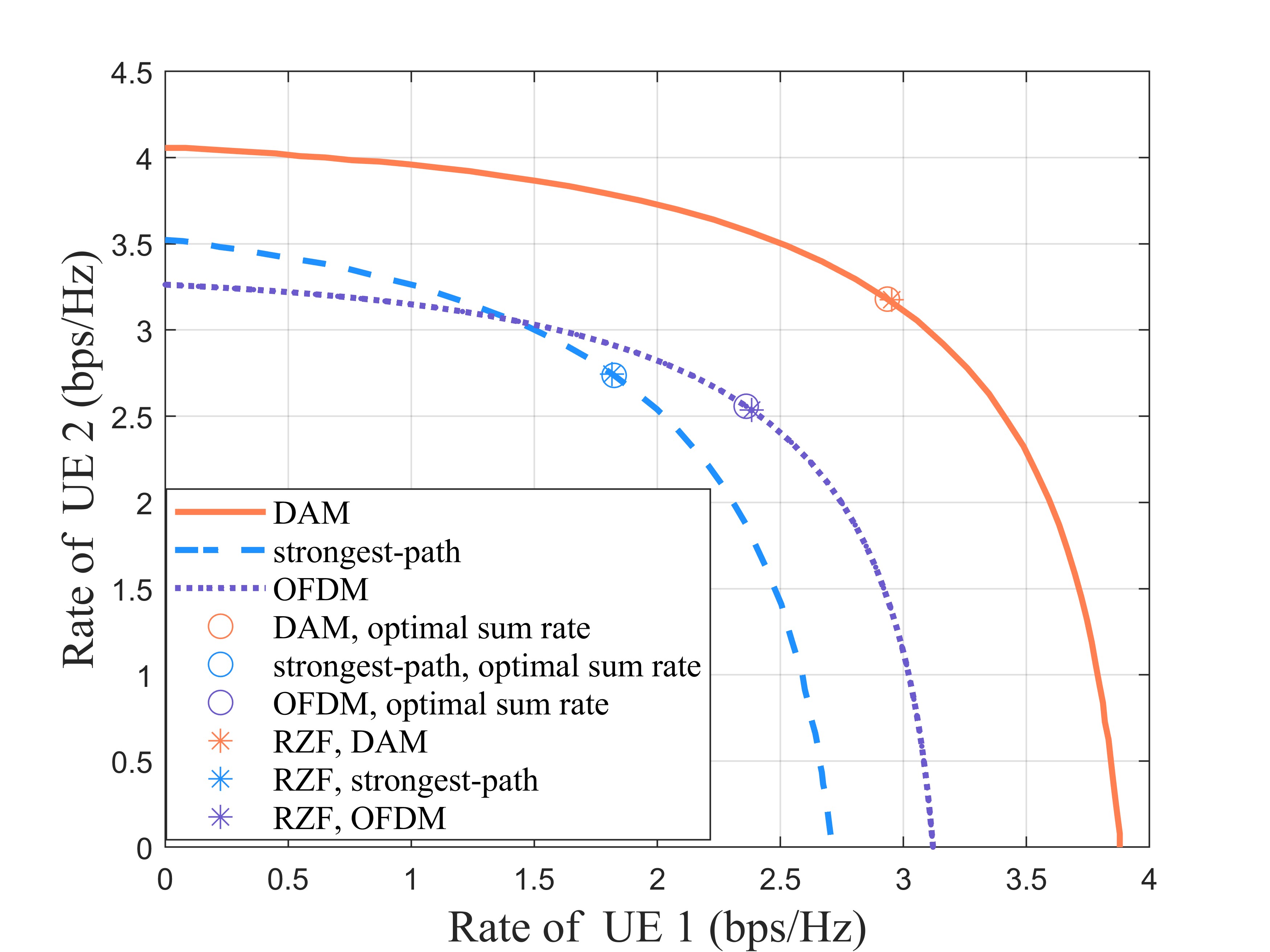}
\caption{Achievable rate region of DAM, strongest-path beamforming, and OFDM.}\label{fig3}
\end{figure}

Fig. \ref{fig3} shows the Pareto boundary for  multi-user DAM, the  strongest-path-based beamforming and OFDM  after considering  the CP or the  guard interval overhead. It is observed  that  the rate region of multi-user DAM is significantly larger than that of the two benchmarking schemes, thanks to its  full utilization of all multi-path components and the low guard interval overhead. 
Besides, the performance of the RZF beamforming scheme approaches to the optimal sum rate on the pareto boundary, which demonstrates the superiority of the RZF beamforming.

\begin{figure}[t]
\centering
	\includegraphics[scale=0.06]{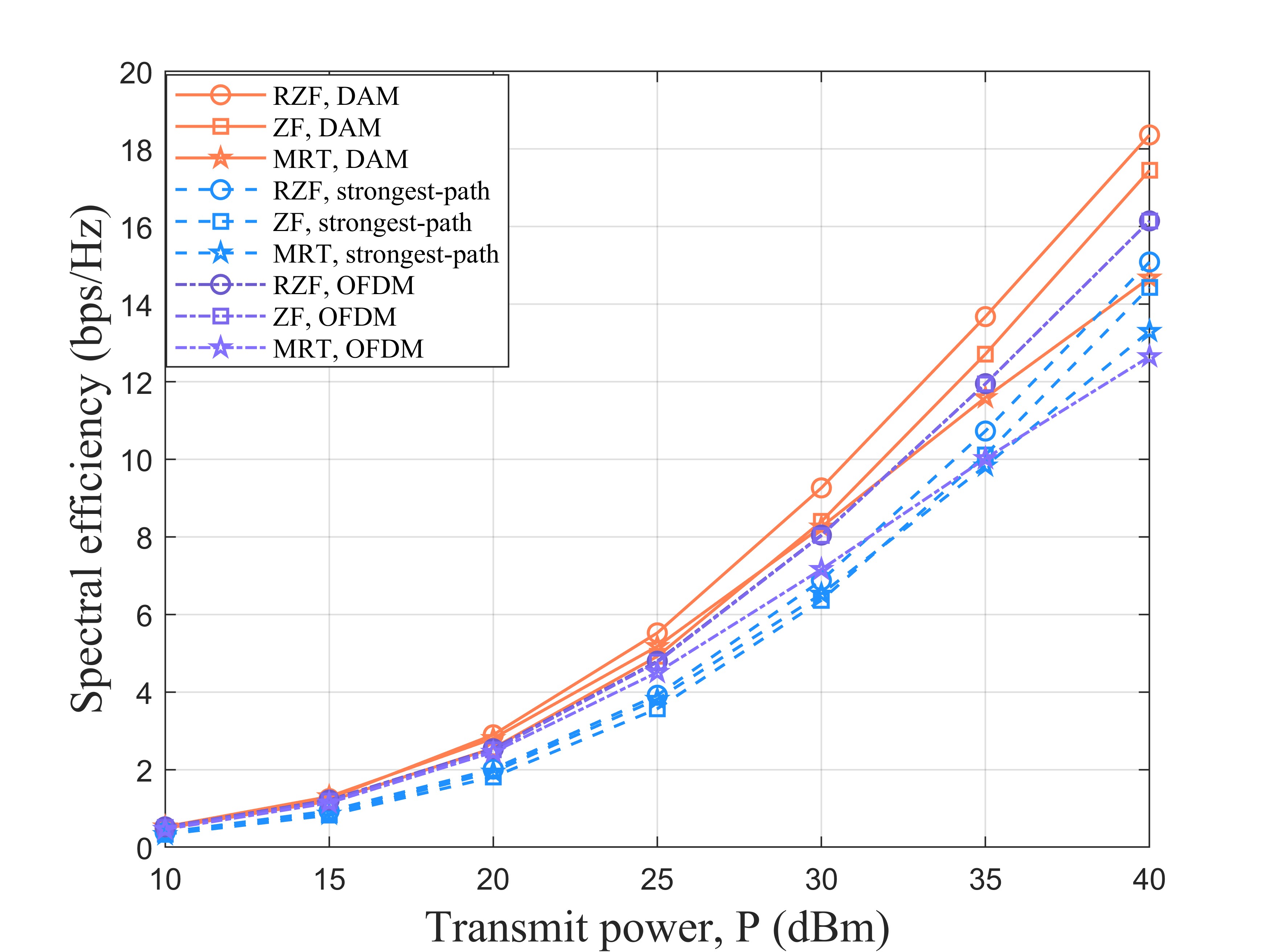}
\caption{Spectral efficiency versus transmit power for the proposed multi-user DAM transmission,  the  strongest-path-based beamforming,  and OFDM.}\label{fig5}
\end{figure}

Fig. \ref{fig5} shows the spectral efficiency versus the transmit  power for the proposed  multi-user DAM  transmission, the  strongest-path-based beamforming and OFDM transmission, with MRT, ZF and RZF beamforming, respectively. It is observed that the proposed multi-user DAM transmission significantly outperforms the two benchmarking schemes  for all the three beamforming strategies. This is expected since DAM makes full use  of the multi-path signal components, as can be seen from the first term in \eqref{4}, whereas the strongest-path-based beamforming only uses the strongest multi-path channel component as the desired signal. On the other hand, DAM is superior to OFDM since DAM requires less guard interval overhead than OFDM, as can be seen in  \eqref{GI-1} and \eqref{GI-2}. It is also observed that  the low-complexity per-path based MRT beamforming achieves comparable performance as ZF and RZF schemes, especially in the low-power regime, thanks to the superior spatial resolution and multi-path sparsity of mmWave massive MIMO systems.

\begin{figure}[t]
\centering
	\includegraphics[scale=0.06]{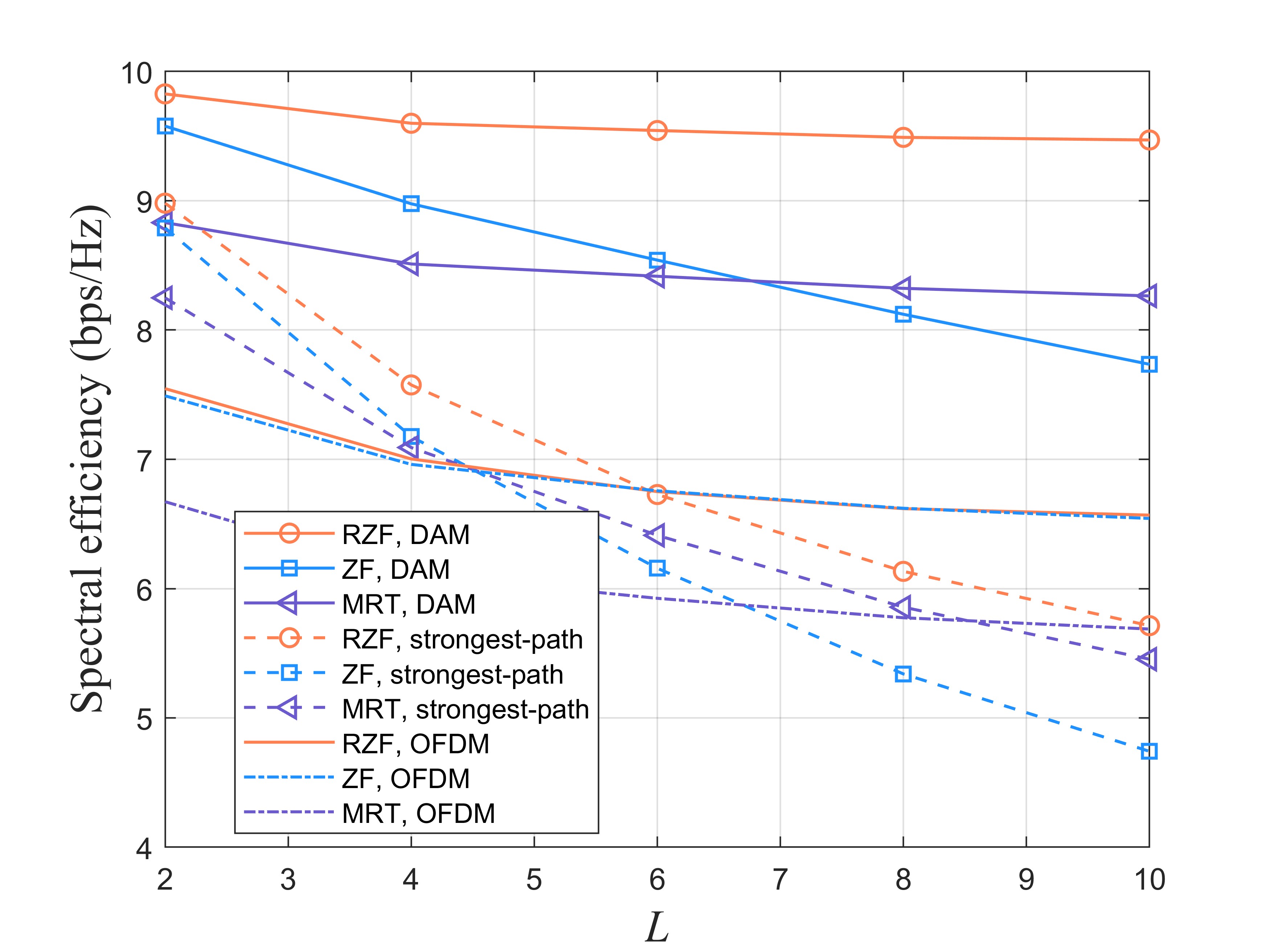}
\caption{Spectral efficiency versus the number of multi-paths for the proposed multi-user DAM and the benchmarking strongest  path  scheme and OFDM transmission.}\label{fig6}
\end{figure}

Fig. \ref{fig6} studies the impact of the number of  multi-paths on the spectral efficiency for the  three schemes. It is observed that the proposed  multi-user  DAM transmission yields a better spectral efficiency performance  than the benchmarking schemes, and both DAM transmission and OFDM show robustness to the increase of the multi-paths. By contrast,  the performance of the  strongest-path-based beamforming degrades significantly with the number of multi-paths. This is because DAM benefits from  all the multi-paths components and OFDM transmits signals with parallel sub-carriers, while the  strongest-path-based beamforming only uses the strongest multi-path channel component.

\begin{figure}[t]
\centering
	\includegraphics[scale=0.06]{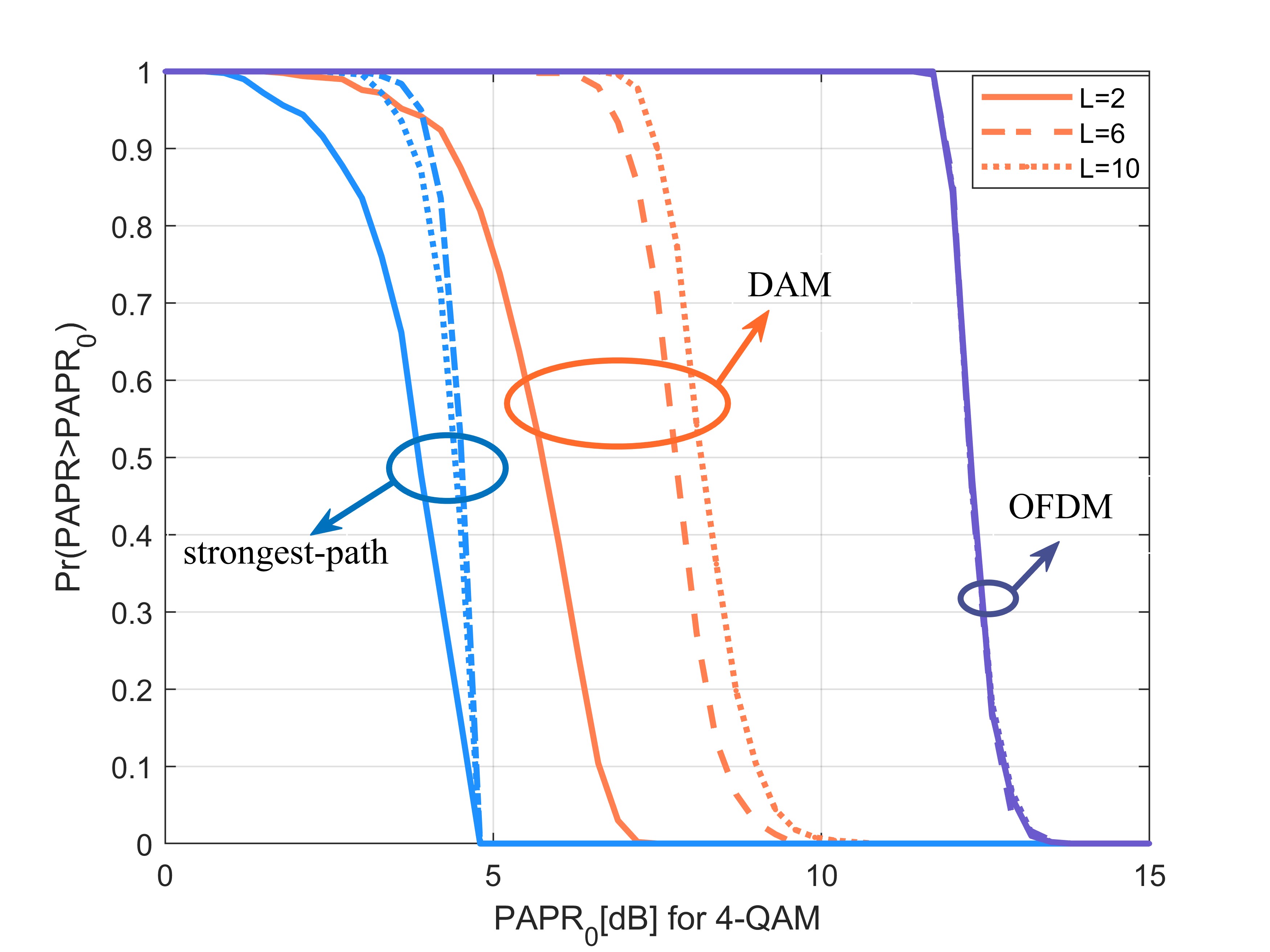}
\caption{PAPR comparison for DAM transmission, OFDM, and  the strongest-path-based scheme  with 4-QAM modulation.}\label{fig7}
\end{figure} 

Last, Fig. \ref{fig7} shows the PAPR comparison for the proposed multi-user DAM transmission, the  strongest-path-based beamforming and OFDM with the corresponding MRT beamforming strategies, by adopting with 4-QAM. Specifically, the metric of complementary cumulative distribution function (CCDF)  is considered to evaluate the PAPR performance. It is observed   that the single-carrier DAM technique achieves significantly lower PAPR than OFDM. This is expected since for OFDM, a total of $KM$ singals  are superimposed  on each antenna, whereas for DAM, only $L_{\mathrm{tot}}$  multi-path signals are mixed  on each antenna, as can be inferred from \eqref{PAPR-6} and \eqref{PAPR-3}, respectively. On the other hand, the PAPR of DAM is slightly higher than the strongest-path-based beamforming, since the strongest-path-based beamforming only uses the single strongest path component as the desired signal, and thus $K$ signals are mixed on each antenna.

\section{Conclusion}
This paper investigated the performance of the multi-user DAM technique for  mmWave massive MIMO communications. For the asymptotic case when the number of BS antennas is much larger than the total number of channel multi-paths, it was shown that both the ISI and IUI are completely eliminated with the simple per-path-based MRT beamforming and delay pre-compensation. For the general scenario with a finite number of BS antennas, the optimal achievable rate region for multi-user  DAM system and benchmarking schemes were characterized. Next,  three classical beamforming schemes   were tailored  for multi-user DAM communication in a per-path basis and benchmarking schemes, so as  to maximize the sum rate, namely the per-path-based MRT, ZF and RZF beamforming. Moreover, the guard interval overhead and PAPR were analyzed for multi-user DAM and the two benchmarking schemes.  Simulation results demonstrated that the proposed multi-user DAM transmission  outperforms the benchmarking schemes of the strongest-path-based beamforming and OFDM  in terms of spectral efficiency, and has a lower PAPR  than OFDM.

%\begin{thebibliography}{1}
\footnotesize
\bibliographystyle{IEEEtran}
\bibliography{Achievable_Rate_Region_and_Path-Based_Beamforming_for_Multi-User_Single-Carrier_Delay_Alignment_Modulation}
%\begin{thebibliography}{1}

\end{document}